\documentclass[journal]{IEEEtran}

\usepackage{array,graphicx,cite,epsfig,amssymb,amsfonts}
\usepackage{bbm,amsmath,epstopdf,algorithm,algorithmic,subfigure, multirow}
\usepackage{comment}
\usepackage{color}
\usepackage{cases}
\usepackage{slashbox}

\newcommand{\be}{\begin{equation}}
\newcommand{\ee}{\end{equation}}
\def\bee#1\eee{\begin{align}#1\end{align}}
\newcommand{\bse}{\begin{subequations}}
\newcommand{\ese}{\end{subequations}}
\newcommand{\nnb}{\nonumber}

\newtheorem{theorem}{\textbf{Theorem}}

\newtheorem{lemma}{\textbf{Lemma}}

\newcommand{\specialcell}[2][c]{%
  \begin{tabular}[#1]{@{}c@{}}#2\end{tabular}}

\newcommand{\red}[1]{{\color{red}#1}}

\newcommand{\bm}[1]{\boldsymbol{#1}}

\begin{document}


\title{\LARGE \bf
On Stability Condition of Wireless Networked Control Systems under Joint Design of Control Policy
and Network Scheduling Policy
}

\author{Lei Deng,~\IEEEmembership{Member,~IEEE},
        Cheng Tan,~\IEEEmembership{Member,~IEEE},
        and Wing Shing Wong,~\IEEEmembership{Fellow,~IEEE}
\thanks{This work was partially funded by Schneider Electric, Lenovo Group (China) Limited and the Hong Kong Innovation and Technology Fund (ITS/066/17FP) under the HKUST-MIT Research Alliance Consortium, and was also partially funded by a grant from the Research Grants Council of the Hong Kong Special Administrative Region under Project GRF 14200217,
the National Natural Science Foundation of China under Grants 61803224.
A preliminary version of this paper has been accepted in IEEE CDC 2018 \cite{deng2018stability}.}
\thanks{L. Deng is with School of Electrical Engineering \& Intelligentization, Dongguan University of Technology and
with Department of Information Engineering, The Chinese University of Hong Kong
        (e-mail: denglei@dgut.edu.cn).}%
\thanks{C. Tan is with College of Engineering, Qufu Normal University and
Department of Information Engineering, The Chinese University of Hong Kong
        (e-mail: tancheng1987love@163.com).}%
\thanks{W. Wong is with Department of Information Engineering, The Chinese University of Hong Kong
        (e-mail: wswong@ie.cuhk.edu.hk).}%
}

\maketitle
\thispagestyle{empty}
\pagestyle{empty}

\begin{abstract}
In this paper, we study a wireless networked control system (WNCS) with $N \ge 2$ sub-systems sharing a common wireless channel.
Each sub-system consists of a plant and a controller and the control message must be delivered from the
controller to the plant through the shared wireless channel. The wireless channel is unreliable due to interference and fading.
As a result, a packet can be successfully delivered in a slot with a certain probability.
A network scheduling policy determines how to transmit those control messages generated by such $N$ sub-systems and
directly influences the transmission delay of control messages.
We first consider the case that all sub-systems have the same sampling period.
We characterize the stability condition of such a WNCS under the joint design of the
control policy and the network scheduling policy by means of $2^N$ linear inequalities.
We further simplify the stability condition into only one linear inequality for two special cases:
the perfect-channel case where the wireless channel can successfully deliver
a control message with certainty in each slot, and the symmetric-structure case where all sub-systems
have identical system parameters.
We then consider the case that different sub-systems can have different sampling periods,
where we characterize a sufficient condition for stability.
\end{abstract}


\section{Introduction} \label{sec:introduction}
Networked Control Systems (NCSs) that exchange information between a plant and its controller through a shared communication network
have been a topic of active research for decades in both the academia and the industry \cite{gupta2010networked,baillieul2007control,hespanha2007survey}.
Existing communication networks employed in NCS include controller-area network (CAN), Ethernet, and wireless networks
(called wireless NCS (WNCS)) \cite{gupta2010networked}.
Among them, WNCS  is widely used in many applications such as automated highway systems, factories, and unmanned aerial vehicles (UAVs), etc.,
because wireless communication can be easily deployed with low cost and low complexity \cite{gupta2010networked, al2016design, park2017wireless,hui2012stabilizing}.
In this paper, we focus on systems based on WNCS.

Traditional NCS researches typically focus on control-theoretic issues while  highly abstracting the network-system performance
in terms of transmission delay and packet dropout/loss \cite{demirel2014modular}.
For example, the behavior of a wireless communication network with finite capacity is commonly modeled by assuming that  packets traveling through the shared communication network
experience a fixed or random delay (with a known distribution) \cite{gao2008new}
or by a fixed packet dropout rate \cite{hu2007}. Using types of simplifications, many researches focus on
how to design the control policy so as to stabilize the system \cite{tan2015stabilization, lam2007} or optimize the system performance \cite{chen2016}.
In \cite{tan2017TAC}, the authors consider  stabilization problems for  NCSs with both packet dropout and transmission delay.
By utilizing a delay-dependent algebraic Riccati equation, a necessary and sufficient stabilization condition is derived.

However, packet delay and packet dropout incurred in a shared communication
network are results of network operations as defined by network protocol or network scheduling policy.
To completely understand the behaviour and performance of NCSs,
it is important to simultaneously consider both the control policy in the dynamic system and
also the network scheduling policy in the network system. There are some existing works that consider such
joint design for WNCS \cite{liu2004wireless,park2011wireless,demirel2014modular} (also see a survey  \cite{park2017wireless} and the references therein).
Reference \cite{demirel2014modular} considers a WNCS with only one plant and one controller which exchanges
information through  a multi-hop wireless network. The authors jointly design the control policy and
network scheduling policy to minimize the closed-loop loss function and propose a modular co-design framework to
solve the problem.  The authors in \cite{park2011wireless}
analyze the system performance of WNCS with multiple plants and controllers when the wireless communication network adopts the standard IEEE 802.15.4 protocol.
The authors in \cite{liu2004wireless} also study a WNCS with multiple plants and controllers and they
analyze the system performance by jointly considering control policy and cross-layer network design.
However, to the best of our knowledge, there does not exist work on characterizing the stability condition of a WNCS with multiple plants and controllers sharing a common wireless channel.

We also remark that the joint design of control system and network system also influences the network scheduling policy design.
Generally, the central performance metric of network system design is throughput in the delay-unconstrained case
or timely throughput in the delay-constrained case. However, a control system may not only depend on
the long-term throughput or timely throughput, but may also depend on the sampled paths, i.e., the short-term behaviors (see more details in later analysis),
which also poses more challenges to the network system design.

In this paper, we study a WNCS with multiple sub-systems sharing a common wireless channel.
Each sub-system consists of a plant and a controller and the control message must be delivered from the
controller to the plant through the shared wireless channel. We
characterize the stability condition of such a WNCS under the joint design of the control policy and the network scheduling policy. In particular, we make the following contributions,
and summarize the main results on stability condition  in Table \ref{tab:resutls-stability condition}:

\begin{itemize}
\item For the stated WNCS with general system parameters so that all sub-systems could have
different parameters (asymmetric-structure case in Table \ref{tab:resutls-stability condition}) and the
wireless channel could be imperfect (imperfect-channel case in Table \ref{tab:resutls-stability condition}),
we characterize the stability condition by means of $2^N$ linear inequalities, where $N$ is the number of sub-systems.
\item We simplify the general stability condition into only one linear inequality for two special cases of the considered WNCS:
the perfect-channel case and the symmetric-structure case.
\item For perfect-channel case, we show that the system can be stabilized if the sampling period is larger than a threshold.
This result quantifies the effect of the sampling period on both the network system and the control system.
\item We show a monotonic property of the stability region in terms of the wireless channel quality:
if the WNCS can be stabilized under a channel quality vector, it can also be stabilized under
a better channel quality vector. This result enables us to efficiently find the minimum channel quality
to stabilize the system for the special symmetric-structure case.
\item Our previous analysis assumes that all sub-systems have the same sampling period.
With this assumption removed, we also characterize a sufficient condition for stability under the case of heterogenous  sampling periods.
\end{itemize}

The rest of this paper is organized as follows.
We first describe our system model in Sec.~\ref{sec:model}.
We next present the stability condition for a general WNCS in Sec.~\ref{sec:stability}. Then, in Sec.~\ref{sec:perfect-channel}
and Sec.~\ref{sec:symmetric-case}, we simplify the general stability condition
into only one linear inequality for two special cases.
In Sec.~\ref{sec:monotonic}, we prove a monotonic result in terms of channel quality.
In addition, we propose a sufficient condition for stability when different sub-systems
have different sampling periods in Sec.~\ref{sec:heterogenous-sampling-periods}. We use simulation to validate our theoretical analysis
in Sec.~\ref{sec:simulation}. Finally, we conclude our paper in Sec.~\ref{sec:conclusion}.
Throughout this paper, we define set $[C] \triangleq \{1,2,\cdots,C\}$
for any positive integer $C\in \mathbb{Z}^+$.

We also remark that  as compared with our preliminary conference version \cite{deng2018stability}, this paper
presents more new results, including (i) the monotonic result in terms of channel quality
in Sec.~\ref{sec:monotonic}, (ii) a sufficient condition for stability when different sub-systems
have different sampling periods in Sec.~\ref{sec:heterogenous-sampling-periods}, and (iii) more simulation results in Sec.~\ref{sec:simulation} (see Fig.~\ref{fig:stability-monotonic-N-p}, Fig.~\ref{fig:min-p-A}, and
Fig.~\ref{fig:min-p-N}).

\begin{table}[t]
\centering
\caption{Main results on stability condition}
\label{tab:resutls-stability condition}
\begin{tabular}{|c|c|c|}
\hline
                    & Symmetric Structure & Asymmetric Structure \\ \hline
\specialcell{Perfect \\ Channel}     & \specialcell{\eqref{equ:stability-condition-symmetric-and-perfect-simplify-h-le-N} for $h < N$, one inequality \\ \eqref{equ:stability-condition-symmetric-and-perfect-simplify-h-g-N} for $h\ge N$, one inequality}                   & \eqref{equ:stablity-condition-perfect-channel}, one inequality                       \\ \hline
\specialcell{Imperfect \\ Channel} & \specialcell{\eqref{equ:stability-condition-symmetric-simplify-h-le-N} for $h < N$, one inequality \\ \eqref{equ:stability-condition-symmetric-simplify-h-g-N} for $h \ge N$, one inequality}                  & \eqref{equ:stability-condition-capacity-region}, $2^N$ inequalities                      \\ \hline
\end{tabular}
\end{table}

\begin{figure}
  \centering
  \includegraphics[width=0.95\linewidth]{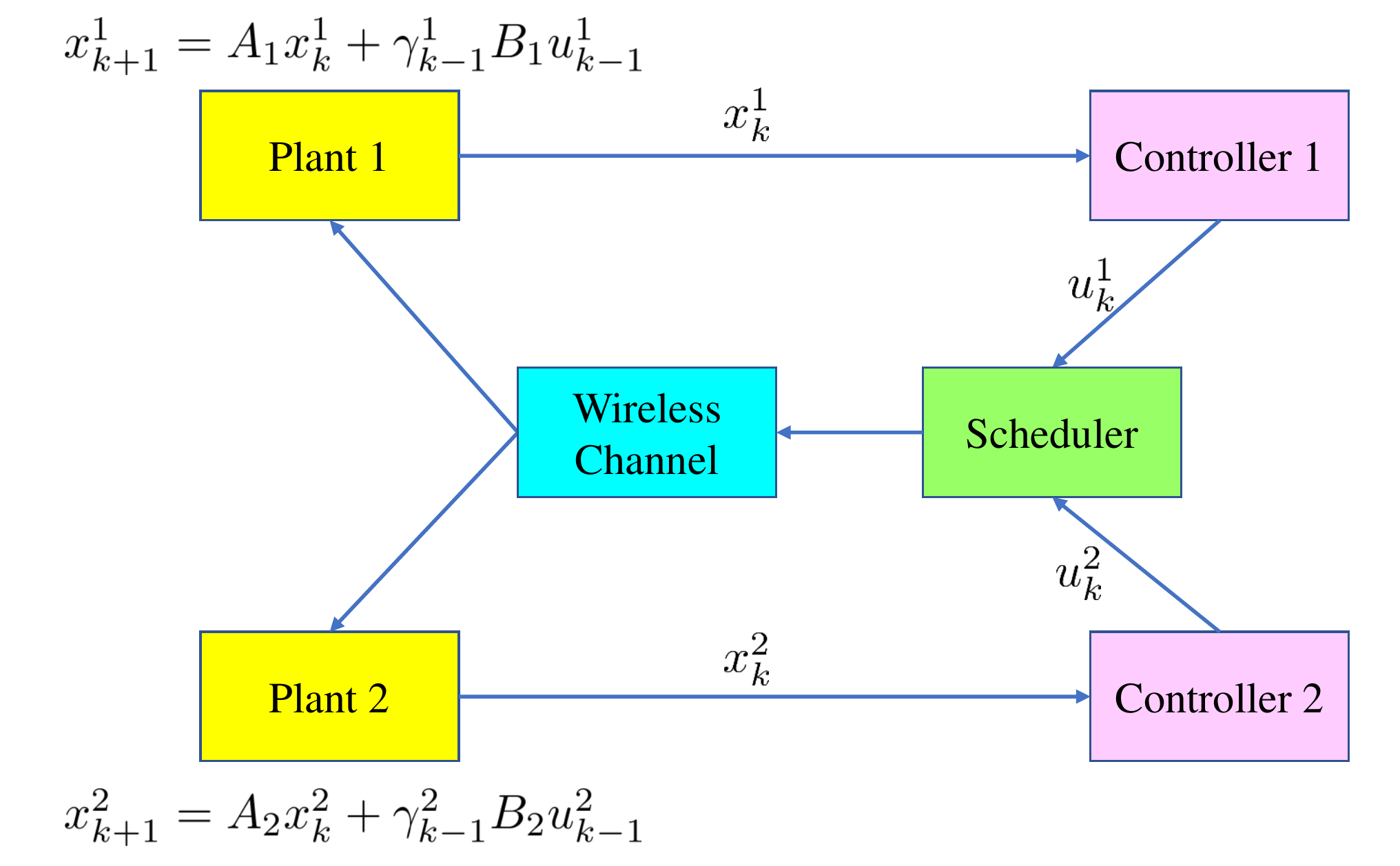}\\
  \caption{System model --- an example of two sub-systems (i.e., $N=2$).}\label{fig:model}
\end{figure}

\section{System Model} \label{sec:model}
We consider a WNCS with  $N$ sub-systems, indexed from 1 to $N$.
An example of two sub-systems is shown in Fig.~\ref{fig:model}. Sub-system $i \in [N]$ has a plant (plant $i$) and a controller (controller $i$). Each plant has
a sensor and an actuator. The sensor can sample and transmit their measurements to the controller over
a dedicated channel without incurring packet loss and delay. The controller makes control decision
based on  sensor's measurements. The control message/packet of the controller is transmitted to the
actuator of the plant to influence the dynamics of the plant over a shared wireless channel, which could incur
packet loss and delay.
A typical practical scenario of our model is the remote control of a fleet of unmanned aerial vehicles (UAVs) \cite{zeyu2016autonomous}.

\begin{figure}
  \centering
  \includegraphics[width=0.95\linewidth]{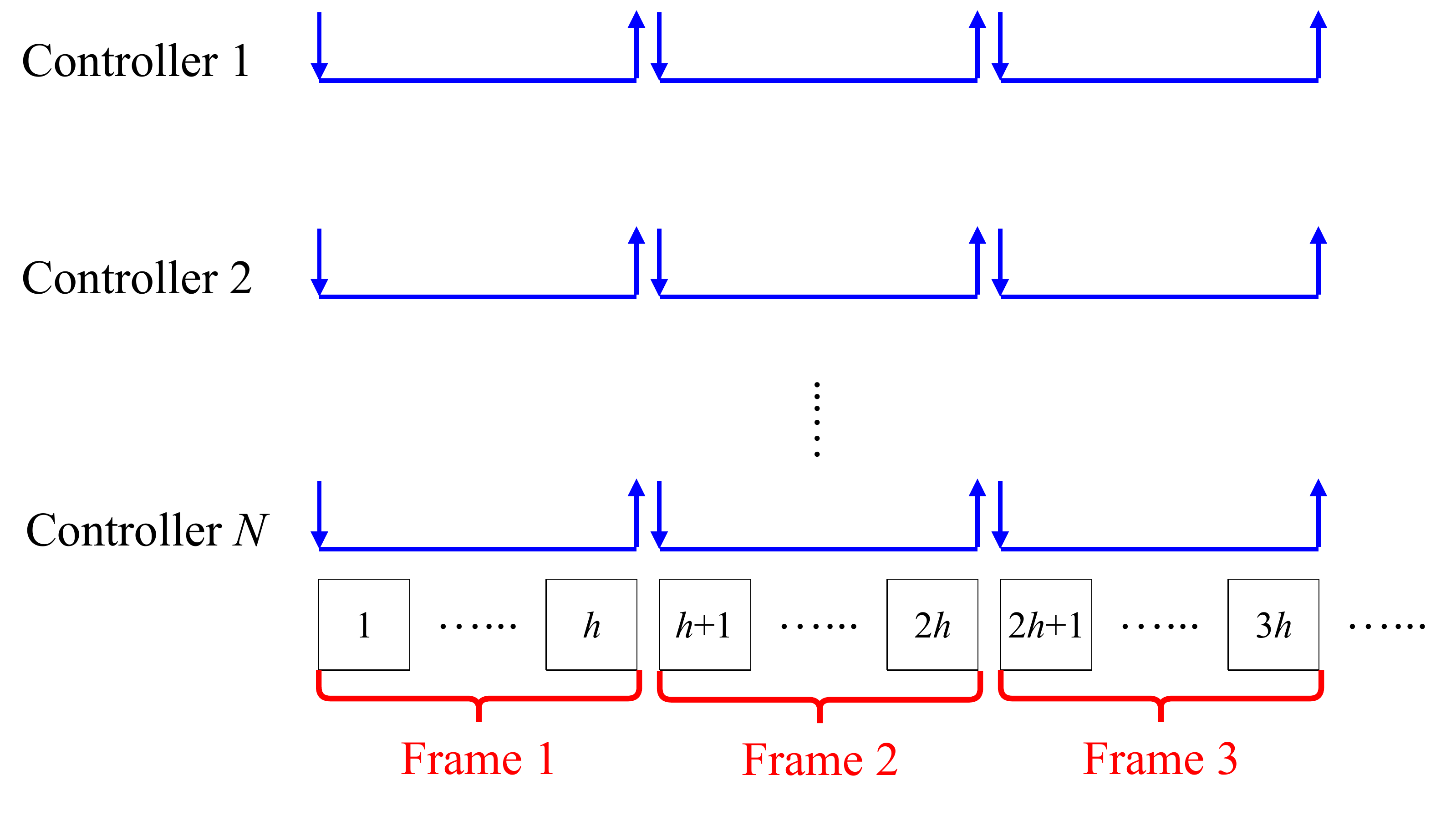}\\
  \caption{Control message/packet pattern.}\label{fig:traffic}
\end{figure}

\textbf{Sub-System Dynamics.}
The underlaying time system is continuous (starting from the initial time $t_0=0$)
but we also create a slotted model (starting from slot 1) for the wireless transmission model
where each slot spans $\Delta>0$ units of time. Details of wireless transmission model will be provided below.
Plant $i$'s underlaying state evolves according to the following continuous-time system:
\be
\dot{x}^i(t) = A_i x^i(t) + B_i u^i(t),\label{sys001}
\ee
where $x^i(t) \in \mathbb{R}$ is the state and $u^i(t) \in \mathbb{R}$ is the control input at time $t$.
To guarantee that each sub-system is stabilizable, we assume that $A_i \ge 0$, and $B_i \neq 0$.\footnote{In this paper, we consider the scalar-state case. It is interesting
and important to extend our results to the general vector-state case.}
Moreover, we assume that each sensor samples the state at the beginning of every $h \in \mathbb{Z}^+$ slots (i.e., every $h\Delta$ units of time).
Namely, we observe plant $i$ every $h$ slots.
Starting from slot 1,
every $T$ slots forms a frame, as shown in Fig.~\ref{fig:traffic}.
For example, frame $k$ is from slot $(k-1)h+1$ to slot $kh$. We observe plant $i$ at the
beginning of each frame $k$ (i.e., at slot $(k-1)h+1$), which is denoted as $x_k^i \in \mathbb{R}$.
Controller $i$ can instantaneously obtain plant $i$'s state $x^k_i $ and then makes a control decision $u_k^i \in \mathbb{R}$. The control
message/packet $u_k^i$ needs to be transmitted to plant $i$ through a shared wireless channel. If $u_k^i$
cannot be delivered before or at the end of frame $k$ (i.e., slot $kh$), a packet dropout occurs. Denote by $\gamma_k^i$  the random variable
which is 1 if message $u_k^i$ is delivered before/at the end of frame $k$ and 0 otherwise.
Then based on the analysis in \cite{cheng2017integrated}, the sampled state of plant $i$ evolves
according to the following discrete-time stochastic system:
\be
x^i_{k+1}=\bar{A}_i x^i_k + \gamma^i_{k-1} \bar{B}_i u^i_{k-1}, \quad k=1,2,\cdots
\label{equ:system-evolution}
\ee
where $\bar{A}_i = e^{A_i h \Delta} \ge 1$ and $\bar{B}_i = \int_{0}^{h \Delta} e^{A_i \tau} B_i d \tau \neq 0$.

Sub-system $i$ is (mean-square) stable if for any initial conditions $x^i_{1}$, $u^i_{0}$ and $\gamma^i_{0}$, the state $x^i_k$ satisfies
\be
\lim_{k \to \infty} \mathbb{E} \left[ |x^i_k|^2 \right] = 0.
\ee
Our goal is to design a stabilizing controller $\{u_k^i:  i \in [N], k=1,2,\cdots\}$ to make all $N$ sub-systems mean-square stable.

\textbf{Wireless Channel and Scheduler.}
The wireless channel is shared by all sub-systems and there is a centralized scheduler
to collect the control packets and then make scheduling decision to transmit them in some order over the wireless channel.
We assume that only one packet can be transmitted over the wireless channel in each slot.
As we mentioned before, each slot spans $\Delta$ units of time, which is the time length
of transmitting a control packet from the scheduler to the plant
and getting the acknowledge about whether the packet is delivered or not
from the plant to the scheduler.

Wireless channel is usually unreliable because of interference and fading. We model such unreliability by a successful probability $p_i \in (0,1]$.
Namely, in a slot, if we transmit the control packet of sub-system $i$,
the message will be delivered successfully with probability $p_i$.\footnote{If a message is delivered successfully, both the
control packet and the acknowledgement packet have successfully reached their destinations.}
Different plants are served with different channel quality depending on their distances  from the scheduler and
different ambient conditions. Thus, the successful probability $p_i$ varies over sub-system index $i$.

\textbf{Design Spaces.}
To make all sub-systems stable, our design spaces include two parts:
\begin{itemize}
\item \emph{The control policy\footnote{We consider linear control policies with respect to the predictive state in this paper \cite{tan2015stabilization}.}}
    $\{u_k^i: i \in [N], k=1,2,\cdots \}$, which determines the control variable $u_k^i$ for each frame $k$ and each sub-system $i$;
\item \emph{The scheduling policy}, which determines the packet to transmit  at each slot.
Note that the distribution of random variable $\gamma_k^i$ is completely determined by the scheduling policy.
\end{itemize}
Both the control policy and the scheduling policy influence the dynamics of the plants according
to equation \eqref{equ:system-evolution}.

\textbf{Assumption on Scheduling Policy.}
In principle, for any sub-system $i$, the joint distribution of random variables $\{\gamma_k^i: k=1,2,\cdots\}$ can
be completely arbitrary because the scheduling policy is arbitrary. However,
it would be difficult to design control policy to stabilize system \eqref{equ:system-evolution} in the mean square sense when the joint distribution of random variables $\{\gamma_k^i: k=1,2,\cdots\}$ has no pattern.
To the best of our knowledge, current literature on NCS only analyzes the case that $\{\gamma_k^i: k=1,2,\cdots\}$ are identical and independent distributed (i.i.d.) (see \cite{tan2015stabilization}).
Therefore, to judiciously leverage the existing results on NCS and delay-constrained wireless communication (see our analysis in the next section),
we only consider the set of scheduling policies such that $\{\gamma_k^i: k=1,2,\cdots\}$ are i.i.d.
We call them \emph{i.i.d. scheduling policies}.

\section{Stability Analysis} \label{sec:stability}
In this section, we leverage existing results in control system as shown in Sec.~\ref{subsec:max-dropout-rate}
and delay-constrained wireless communication as shown in Sec.~\ref{subsec:timely-capacity-region}
to characterize the stability region of our considered WNCS in Sec.~\ref{subsec:stability}.

\subsection{Maximum Dropout Rate of Control System \cite{tan2015stabilization}} \label{subsec:max-dropout-rate}

According to \cite[Theorem 3]{tan2015stabilization}, for any sub-system $i$, if $\{\gamma^i_{k}: k=1,2,\cdots,\}$ are i.i.d. with ${P}(\gamma^i_k=0) = q_i$, where $q_i \in [0,1]$ is called the packet dropout rate of sub-system $i$, then
sub-system $i$ is (mean-square) stable if and only if\footnote{For the scalar case, the stability of sub-system $i$ does not depend on parameter $B_i$. The reason is that
the control policy can determine the control variables $\{u^i_k\}$ to compensate the effect of parameter $B_i$ (see system dynamics \eqref{sys001} and \eqref{equ:system-evolution}).
}
\be
q_i < q^i_{\max}\left({A}_i, h\right) \triangleq \frac{1}{\bar{A}_i^4 - \bar{A}_i^2 + 1} = \frac{1}{ e^{4A_ih \Delta} - e^{2A_ih \Delta} + 1}. \label{equ:max-dropout-rate}
\ee

Here $q^i_{\max}\left({A}_i, h\right)$ is the maximum dropout rate of sub-system $i$, which depends on parameter $A_i$ and sampling period $h$.\footnote{
In fact,  $q^i_{\max}\left({A}_i, h\right)$ defined in \eqref{equ:max-dropout-rate} also depends on slot length $\Delta$. However,
for simplicity, we ignore it in the notation of $q^i_{\max}\left({A}_i, h\right)$ since we do not evaluate the effect of slot length $\Delta$.}
Sub-system $i$ can be stabilized if and only if its dropout rate $q_i$ is strictly less than the maximum dropout rate $q^i_{\max}\left({A}_i, h\right)$.
Clearly, any i.i.d. scheduling policy induces a dropout rate $q_i$ for any sub-system $i$.
Thus to stabilize all sub-systems, we need to design an i.i.d. scheduling policy such that
\eqref{equ:max-dropout-rate} holds for any sub-system $i \in [N]$.

\textbf{Control Policy Design.}
For given dropout rate $q_i={P}(\gamma_k^i=0)$, we only mentioned the stability condition \eqref{equ:max-dropout-rate} of sub-system $i$ but ignore the design of control policy.
In fact, if $q_i$ satisfies \eqref{equ:max-dropout-rate}, we can solve the following delay-dependent algebraic Riccati equation (DARE) whose variable is $P_i>0$,
\begin{equation}
\label{dare01}
P_i=\bar{A}_i' P_i\bar{A}_i+I-M_i'\Upsilon_i^{-1}M_i,
\end{equation}
with
\begin{align}
\Upsilon_i=&(1-q_i)^2 \bar{B}_i'P_i \bar{B}_i+q_i(1-q_i)\bar{B}_i'\bar{A}_i' P_i \bar{A}_i \bar{B}_i \nonumber \\
&+q_i(1-q_i)\bar{B}_i'\bar{B}_i+I, \nonumber \\
M_i=&(1-q_i)\bar{B}_i'P_i \bar{A}_i. \label{gare02}
\end{align}
Note that \eqref{dare01} and \eqref{gare02} are for general vector-state case. For our scalar-state case,
we can simplify them by applying $\bar{A}'_i=\bar{A}_i$, $\bar{B}'_i=\bar{B}_i$, ${M}'_i={M}_i$, and $I=1$.
Moreover, the stabilizing control policy is designed as
\bee
u^i_{k} & =-\Upsilon_i^{-1}M_i \hat{x}^i_{k+1|k-1} ,
\label{equ:control-policy}
\eee
where
\be
\hat{x}^i_{k+1|k-1} \triangleq \bar{A}_i x^i_{k} + (1-q_i)\bar{B}_i u^i_{k-1}
\label{equ:predicted-state}
\ee
is the predicted state in frame $k+1$ based on the observation of
the state in frame $k$ and the control variable in frame $k-1$.
Please refer to \cite{tan2015stabilization} for the details and proofs.

\subsection{Timely Capacity Region of Network System \cite{hou2009qos,hou2012queueing,deng2017timely}} \label{subsec:timely-capacity-region}
In our network system, each packet has a hard deadline of $h$ slots and it becomes
useless if it cannot be delivered before its deadline.
The major performance metric of such delay-constrained communication is \emph{timely throughput}.
In particular, the timely throughput
of sub-system $i$ is the long-term per-frame average number of packets that are successfully delivered before their deadlines \cite{hou2009qos,hou2012queueing,deng2017timely},
i.e.,
\be \label{equ:def-timely-throughput}
R_i \triangleq \liminf_{K \to \infty} \frac{\mathbb{E}[\sum_{k=1}^{K} \gamma_k^i] }{K},
\ee
which depends on the scheduling policy.
The (timely) capacity region
is the set of timely throughput vector $\bm{R}=(R_1,R_2,\cdots,R_N)$ such that there exists
a scheduling policy under which the sub-system $i$'s timely
throughput is at least $R_i$ for any $i \in [N]$.
Note that when we characterize the capacity region, we consider
all possible scheduling polices (not necessarily i.i.d.).
In addition, since the capacity region depends on
both the channel quality vector $\bm{p} \triangleq (p_1,p_2,\cdots,p_N)$
and the frame length (i.e. sampling period)
$h$, we denote it as $\mathcal{R}(\bm{p}, h)$.

In the literature on delay-constrained wireless communication, there are two equivalent characterizations for the capacity region $\mathcal{R}(\bm{p}, h)$:
one idle-time-based in \cite{hou2009qos,hou2012queueing} and one MDP-based in \cite{deng2017timely}.

\textbf{Idle-Time-Based Approach.}
Denote $\Theta_i$ as the number of transmissions until one gets a successful delivery for sub-system $i$'s control message,
which is a geometric random variable with mean $1/p_i$.
Since we have in total $h$ slots in a frame,
the number of \emph{idle slots} in a frame when we only schedule the control packets in sub-system set $\mathcal{S} \subset [N]$ in any work-conserving manner
is
\be
I_{\mathcal{S}} \triangleq \max \left\{h - \sum_{i \in \mathcal{S}} \Theta_i, 0 \right\}.
\label{equ:def-I-S}
\ee
Note that the distribution of  $I_{\mathcal{S}}$ is the same for any work-conserving scheduling policy \cite{hou2009qos}.

Hou \emph{et al.} in \cite{hou2009qos,hou2012queueing} proved that
the capacity region $\mathcal{R}(\bm{p}, h)$ is the set of all timely throughput vectors
$\bm{R}=(R_1, R_2, \cdots, R_N)$ satisfying
\be \label{equ:capacity-region}
\sum_{i \in \mathcal{S}} \frac{R_i}{p_i} + \mathbb{E}[I_{\mathcal{S}}] \le h, \quad \forall \mathcal{S} \subset [N],
\ee
which contains $2^N$ linear inequalities.

Hou \emph{et al.} in \cite{hou2009qos,hou2012queueing} further proposed the largest-deficit-first (LDF)
scheduling policy and proved that LDF is feasibility optimal in the sense that it can achieve any input
feasible timely throughput vector in the capacity region. However, LDF is frame-dependent and thus
the induced $\{\gamma_k^i:k=1,2\cdots\}$ are not i.i.d. Therefore, it cannot be applied to our control system.
This also illustrates that when we jointly consider the network system and control system, we introduce new challenge to design the network scheduling policy.

\textbf{MDP-Based Approach.}
Deng \emph{et al.} in \cite{deng2017timely} proposed another (equivalent) capacity region characterization based on the Markov Decision Process (MDP) theory.
Denote the state of sub-system $i$ at slot $t$ as
\be
S^i_t \triangleq
\left\{
  \begin{array}{ll}
    1, & \hbox{If sub-system $i$ has a packet at slot $t$ ;} \\
    0, & \hbox{Otherwise.}
  \end{array}
\right.
\ee
Then the state of the whole system at slot $t$ is denoted as
\[
\boldsymbol{S}_t \triangleq (S^1_t,S^2_t,\cdots,S^N_t).
\]
The state space (the set of all possible states) is $\mathcal{S} = \{0,1\}^N$.
The action at slot $t$, denoted as $A_t$, is to determine which sub-system's control packet to transmit.
In particular, $A_t=i$ means to transmit sub-system $i$'s control packet at slot $t$.
Then the action space is $\mathcal{A}=\{1,2,\cdots, N\}=[N]$.
The reward function of sub-system $i$ is denoted as
\be
r_i(\boldsymbol{s},a) = p_i \cdot 1_{\{\text{sub-system $i$ has a packet under state $\bm{s}$ and $a=i$}\}}.
\ee
It is also easy to compute the transition probability from state $\bm{s}$ to state $\bm{s}'$
if taking action $a$ at slot $t$, which is denoted as $P_t(\bm{s}'|\bm{s},a)$.

Deng \emph{et al.} in \cite{deng2017timely}  proved that the capacity region $\mathcal{R}(\bm{p},h)$
is characterized by the following linear inequalities,
\bse \label{equ:capacit-region-MDP}
\bee
& \sum_{a \in \mathcal{A}} x_{t+1}(\bm{s}',a) = \sum_{\bm{s} \in \mathcal{S}} \sum_{a \in \mathcal{A}} x_t(\bm{s},a) P_t(\bm{s}' | \bm{s}, a), \nnb \\
& \qquad \qquad \forall \bm{s}' \in \mathcal{S}, t \in [h], \\
& \sum_{a \in \mathcal{A}} x_{1}(\bm{s}',a) = \sum_{\bm{s} \in \mathcal{S}} \sum_{a \in \mathcal{A}} x_h(\bm{s},a) P_t(\bm{s}' | \bm{s}, a), \nnb \\
& \qquad \qquad \forall \bm{s}' \in \mathcal{S}, \\
& R_i \le \sum_{t \in [h]} \sum_{\bm{s} \in \mathcal{S}} \sum_{a \in \mathcal{A}} {x_t(\bm{s},a)r_i(\bm{s},a)}, \quad \forall i \in [N], \\
&  \sum_{\bm{s} \in \mathcal{S}} \sum_{a \in \mathcal{A}} x_t(\bm{s},a) =1, \quad \forall t \in [h],  \\
&  x_t(\bm{s},a) \ge 0, \quad \forall t\in [h], \bm{s} \in \mathcal{S}, a \in \mathcal{A}.
\eee
\ese
The capacity region $\mathcal{R}(\bm{p},h)$ is the set of all possible $\bm{R}=(R_1,R_2,\cdots,R_N)$
where there exists a set of $\{x_{t}(\bm{s},a)\}$  such that
\eqref{equ:capacit-region-MDP} holds. Note that the MDP-based capacity region
characterization \eqref{equ:capacit-region-MDP} has $O(h\cdot N \cdot 2^N)$ linear equalities/inequalites.

As compared to the idle-time-based approach, another main result of the MDP-based approach (see \cite[Theorem 2]{deng2017timely}) is that any feasible timely throughput vector $\bm{R}$ can be achieved by an i.i.d. scheduling policy.
Therefore, non-i.i.d. scheduling policies cannot enlarge the capacity region $\mathcal{R}(\bm{p},h)$. In addition,
Deng \emph{et al.} in \cite{deng2017timely} also designed an i.i.d. scheduling policy such that it can achieve
any input feasible timely throughput vector. More specifically, for any feasible timely throughput vector $\bm{R}$ (i.e., $\bm{R}$ satisfies \eqref{equ:capacity-region}
or \eqref{equ:capacit-region-MDP}), we obtain a set of $\{x_{t}(\bm{s},a)\}$  such that
\eqref{equ:capacit-region-MDP} holds. Then the i.i.d. scheduling policy is
\be \label{equ:RAC-scheduling-policy}
\left\{
  \begin{array}{ll}
    P_{A_t|S_t}(a|s)=\frac{ x_t(s,a)}{\sum_{a' \in\mathcal{A}}x_t(s,a')}, & \hbox{$\forall t \in [h]$;} \\
    P_{A_t|S_t}(a|s)= P_{A_{t-h}|S_{t-h}}(a|s), & \hbox{$\forall t > h$,}
  \end{array}
\right.
\ee
where $P_{A_t|S_t}(a|s)$ is the probability to transmit sub-system $a$'s control packet at slot $t$
conditioning on that the system state is $S_t=s$ at slot $t$.

Since the traffic patten is frame-synchronized, the system state becomes $\bm{s}=(1,1,\cdots,1)$
in the beginning of each frame and the scheduling policy in \eqref{equ:RAC-scheduling-policy} repeats
every frame, the induced $\{\gamma^i_k:k=1,2,\}$ are i.i.d., which can thus be applied to
our control system as discussed in Sec.~\ref{subsec:max-dropout-rate}. In addition,
according to the definition of timely throughput in \eqref{equ:def-timely-throughput}, we can obtain that
the induced dropout rate is
\be
q_i = P(\gamma^i_k=0) = 1 - P(\gamma^i_k=1) = 1 - R_i,
\ee
which relates the timely throughput to the dropout rate.

\subsection{Stability Condition of Our WNCS} \label{subsec:stability}
We now present our main result on stability condition of our WNCS when both
the control policy and the scheduling policy are taken into consideration.

\begin{theorem} \label{thm:general-stabiliyt-condition}
Suppose that we only consider those scheduling policies such that $\{\gamma_k^i:k=1,2,\cdots\}$ are i.i.d.
Then given system parameters $\bm{A}=(A_1, A_2, \cdots, A_N)$, $\bm{B}=(B_1, B_2, \cdots, B_N)$, $\bm{p}=(p_1, p_2, \cdots, p_N)$, sampling period $h$, and slot length $\Delta$,
there exists a control policy and an i.i.d. network scheduling policy to make all sub-systems (mean-square) stable if and only if
\be
\resizebox{.88\hsize}{!}{$\left(1-q_{\max}^1\left({A}_1,h\right), \cdots, 1-q_{\max}^N\left({A}_N,h\right) \right) \in \text{int}\left(\mathcal{R}(\bm{p},h)\right)$},
\label{equ:stablity-condition-org}
\ee
where $\text{int}(\mathcal{S})$ denotes the interior of a set $\mathcal{S}$.
\end{theorem}
\begin{IEEEproof}
\textbf{``If" Part.} If \eqref{equ:stablity-condition-org} holds,
then there exists a timely throughput vector $\bm{R}=(R_1, R_2, \cdots, R_N) \in \mathcal{R}(\bm{p}, h)$ such that
\be
1-q^i_{\max}({A}_i, h) < R_i.
\label{equ:app-if-equ1}
\ee
As we discussed in Sec.~\ref{subsec:timely-capacity-region}, $\bm{R}=(R_1, R_2, \cdots, R_N)$ can be achieved
by an i.i.d. scheduling policy such that
\be
R_i = P(\gamma^i_k = 1) = 1 - P(\gamma^i_k = 0) = 1 - q_i, \forall k=1,2,\cdots.
\label{equ:app-if-equ2}
\ee
Combining \eqref{equ:app-if-equ1} and \eqref{equ:app-if-equ2},
we have
\be
1-q^i_{\max}({A}_i, h) < R_i = 1 - q_i, \nnb
\ee
implying that
\be
q_i < q^i_{\max}({A}_i, h). \nnb
\ee
Thus, any sub-system $i$ can be stabilized according to our discussion in Sec.~\ref{subsec:max-dropout-rate}.

\textbf{``Only If" Part.} If the system can be stabilized under the condition that
$\{\gamma_k^i:k=1,2,\cdots\}$ are i.i.d., then according to our discussion in Sec.~\ref{subsec:max-dropout-rate},
we have
\be
P(\gamma^i_k = 0)  =  q_i < q^i_{\max}({A}_i, h).
\label{equ:app-only-if-equ1}
\ee
The scheduling policy such that $\{\gamma_k^i:k=1,2,\cdots\}$ are i.i.d. with
$P(\gamma^i_k = 0)  =  q_i$ achieves the timely throughput
$R_i = P(\gamma^i_k = 1) = 1 - q_i$ for any sub-system $i$.
Thus,
\bee
\bm{R} & =(R_1,R_2,\cdots, R_N) \nnb \\
& = (1-q_1, 1-q_2, \cdots, 1-q_N) \in \mathcal{R}(\bm{p}, h).
\label{equ:app-only-if-equ2}
\eee
Combining \eqref{equ:app-only-if-equ1} and \eqref{equ:app-only-if-equ2}, we
prove that \eqref{equ:stablity-condition-org} holds.

The proof is completed.
\end{IEEEproof}

Theorem~\ref{thm:general-stabiliyt-condition} ensures that we only need to check
condition \eqref{equ:stablity-condition-org} to determine whether a system can be stabilized.
We have two different capacity region characterizations,
but the idle-time-based one is easy to analyze.
Thus, following from \eqref{equ:capacity-region},
the stability condition \eqref{equ:stablity-condition-org} is equivalent to the following inequalities
\be
\sum_{i \in \mathcal{S}} \frac{1 - q_{\max}^i({A}_i,h)}{p_i} + \mathbb{E}[I_{\mathcal{S}}] < h, \quad \forall \mathcal{S} \subset [N].
\label{equ:stability-condition-capacity-region}
\ee
Note that in \eqref{equ:stability-condition-capacity-region} we have in total $2^N$ linear inequalities.

If \eqref{equ:stability-condition-capacity-region} holds, i.e., the system can be stabilized,
we can input $\bm{R}=(1 - q_{\max}^1({A}_i,h)+\epsilon,1 - q_{\max}^2({A}_i,h)+\epsilon,\cdots,1 - q_{\max}^N({A}_i,h)+\epsilon)$
where $\epsilon>0$ is a sufficiently small positive real number into \eqref{equ:capacit-region-MDP} and then we
obtain  a set of $\{x_{t}(\bm{s},a)\}$, which is further inserted to \eqref{equ:RAC-scheduling-policy} to construct the i.i.d. scheduling policy.\footnote{Though we use
the idle-time-based approach to characterize the stability condition in \eqref{equ:stability-condition-capacity-region}, we need
to use the MDP-based approach to construct the i.i.d. scheduling policy.}
Inserting $q_i = q_{\max}^i({A}_i,h) - \epsilon$ into \eqref{dare01} and \eqref{gare02}, we obtain $\Upsilon_i$ and $M_i$
and thus construct the control policy following from \eqref{equ:control-policy}.
The constructed control policy and i.i.d. scheduling policy make the whole system mean-square stable.

Next we simplify
the stability condition \eqref{equ:stability-condition-capacity-region} for two special cases: the perfect-channel case (Sec.~\ref{sec:perfect-channel})
and the symmetric-structure case (Sec.~\ref{sec:symmetric-case}), both of which
characterize the stability condition by means of only one linear inequality.

\section{The Perfect-Channel Case} \label{sec:perfect-channel}
Consider $p_i=1$ for all $i \in [N]$, i.e., the wireless channel is perfect in the sense that it can successfully
deliver a packet in each slot with certainty. In this case,  we can simplify the stability condition.
Note that $I_{\mathcal{S}} = \max\{h- |\mathcal{S}|, 0\}$ for perfect channel. Thus, \eqref{equ:stability-condition-capacity-region} becomes
\be
\sum_{i \in \mathcal{S}} \left[1 - q_{\max}^i({A}_i,h) \right] + \max\{h- |\mathcal{S}|, 0\} < h, \quad \forall \mathcal{S} \subset [N], \label{equ:capacity-region-perfect-channel}
\ee

If $|\mathcal{S}| \le h$, then $\max\{h- |\mathcal{S}|, 0\}= h - |\mathcal{S}|$ and thus \eqref{equ:capacity-region-perfect-channel} becomes
\be
\sum_{i \in \mathcal{S}} \left[1 - q_{\max}^i({A}_i,h) \right] < |\mathcal{S}|, \quad \forall \mathcal{S} \subset [N],
\ee
which is definitely true since $1 - q_{\max}^i({A}_i,h) < 1, \forall i$.

If $|\mathcal{S}| > h$, then $\max\{h- |\mathcal{S}|, 0\}= 0$ and thus \eqref{equ:capacity-region-perfect-channel} becomes
\be
\sum_{i \in \mathcal{S}} (1 - q_{\max}^i({A}_i,h)) \le h, \quad \forall \mathcal{S} \subset [N]
\label{equ:capacity-region-perfect-channel-S-larger-than-h}
\ee
When $\mathcal{S} = [N] = \{1,2,\cdots,N\}$, \eqref{equ:capacity-region-perfect-channel-S-larger-than-h} becomes
\be
\sum_{i=1}^N (1 - q_{\max}^i({A}_i,h)) \le h,
\ee
which implies \eqref{equ:capacity-region-perfect-channel-S-larger-than-h} for any other
$\mathcal{S} \subset [N]$.
Therefore, in the perfect-channel case, the stability condition \eqref{equ:stability-condition-capacity-region} becomes
\bee
& \sum_{i=1}^N \left[1-q_{\max}^i({A}_i,h) \right] \hspace{-0.8mm}=\hspace{-0.8mm} \sum_{i=1}^N \left[ 1 - \frac{1}{ e^{4A_ih \Delta} - e^{2A_ih\Delta} + 1} \right]
\hspace{-0.6mm} <\hspace{-0.6mm} h.
\label{equ:stablity-condition-perfect-channel}
\eee
Hence, we characterize the stability condition of the WNCS in the perfect-channel case by means of one linear inequality.

We further show a property for the perfect-channel case.
\begin{theorem} \label{the:h-min-perfect}
There exists an $h_{\min} \in [N]$ such that \eqref{equ:stablity-condition-perfect-channel} holds if the sampling period $h \ge h_{\min}$.
\end{theorem}
\begin{IEEEproof}
Please see Appendix~\ref{app:proof-of-h-min-perfect}.
\end{IEEEproof}

Theorem~\ref{the:h-min-perfect} shows that $h \ge h_{\min}$ is a sufficient condition for stability.
Readers may wonder whether it is also necessary. The answer is negative as shown later in Fig.~\ref{fig:stability-perfect-channel-h} in the simulation section.
Note that the sampling period $h$ influences both the control system and the network scheduling system.
On one hand, the maximum dropout rate $q^i_{\max}\left({A}_i, h\right)$ (see \eqref{equ:max-dropout-rate}) decreases as the sampling period $h$ increases, meaning that
it is more difficult to stabilize the control system when the sampling period increases.
On the other hand, $h$ is the frame length of the network system. It is straightforward to prove that
the capacity region $\mathcal{R}(\bm{p},h)$ increases as $h$ increases. Thus, larger sampling period $h$
can increases the delivery probability of a control message.
Overall, the sampling period $h$ balances the intensity of control messages and the delivery chance/quality of
individual control messages. Theorem~\ref{the:h-min-perfect} shows that we prefer larger $h$ from the perspective of
the overall system: when we increase the sampling period $h$ in the perfect-channel case, the benefit of increasing delivery chance/quality of
individual control messages dominates the cost of decreasing the intensity of control messages.

\section{The Symmetric-Structure Case} \label{sec:symmetric-case}
The capacity region $\mathcal{R}(\bm{p},h)$ becomes much more complicated in the imperfect-channel case, i.e., when some $p_i < 1$.
In this section, we analyze a special class of general channels (which could be perfect or imperfect), known as symmetric-structure case, such that
$A_i=A$, $p_i=p$  for all $i\in [N]$. Thus,
\bee
& q^i_{\max}({A}_i,h)  = \frac{1}{ e^{4Ah \Delta} - e^{2Ah \Delta} + 1} \triangleq q_{\max}({A},h),  \forall i \in [N].
\eee
Due to the symmetric structure, if $|\mathcal{S}_1|=|\mathcal{S}_2|$, we have $\mathbb{E}[I_{\mathcal{S}_1}] = \mathbb{E}[I_{\mathcal{S}_2}]$.
Then the stability condition \eqref{equ:stability-condition-capacity-region} becomes
\bse \label{equ:stability-condition-capacity-region-symmetric-1}
\bee
  & \frac{1-q_{\max} ({A}, h)}{p} + \mathbb{E}[I_{\{1\}}] < h, \\
   & \frac{2\left(1-q_{\max} ({A}, h)\right)}{p} + \mathbb{E}[I_{\{1,2\}}]  < h, \\
& \cdots \\
 & \frac{N\left(1-q_{\max} ({A}, h)\right)}{p} + \mathbb{E}[I_{\{1,2, \cdots, N\}}]  < h,
\eee
\ese
which is equivalent to
\bse \label{equ:stability-condition-capacity-region-symmetric-2}
\bee
  & \frac{1-q_{\max} ({A}, h)}{p}  < h - \mathbb{E}[I_{\{1\}}], \\
  & \frac{1-q_{\max} ({A}, h)}{p}  < \frac{h - \mathbb{E}[I_{\{1,2\}}]}{2}, \\
  & \cdots \\
  & \frac{1-q_{\max} ({A}, h)}{p}  < \frac{h - \mathbb{E}[I_{\{1,2, \cdots, N\}}]}{N},
\eee
\ese
We will further simplify \eqref{equ:stability-condition-capacity-region-symmetric-2} based on the following result.
\begin{theorem} \label{thm:sequential-inequality-due-to-submodular}
In  the symmetric-structure case, we have
\be
 h - \mathbb{E}[I_{\{1\}}] \ge \frac{h - \mathbb{E}[I_{\{1,2\}}]}{2} \ge  \cdots \ge \frac{h - \mathbb{E}[I_{\{1,2, \cdots, N\}}]}{N}. \nnb
\ee
\end{theorem}
\begin{IEEEproof}
See Appendix \ref{app:proof-of-sequential-inequality-due-to-submodular}.
\end{IEEEproof}

Theorem~\ref{thm:sequential-inequality-due-to-submodular} shows that the stability condition \eqref{equ:stability-condition-capacity-region-symmetric-2}
can be further simplified into one linear inequality,
\be
\frac{1-q_{\max} ({A}, h)}{p}  < \frac{h - \mathbb{E}[I_{\{1,2, \cdots, N\}}]}{N}.
\label{equ:stability-condition-capacity-region-symmetric-simplify}
\ee

We next show how to calculate $h - \mathbb{E}[I_{\{1,2, \cdots, N\}}]$.
Note that $h - \mathbb{E}[I_{\{1,2, \cdots, N\}}]=\mathbb{E}[h -I_{\{1,2, \cdots, N\}}]$ is the expected number of transmissions in a
frame of $h$ slots when scheduling all sub-systems' control packets. Denote  random variable $X \triangleq h -I_{\{1,2, \cdots, N\}}$. Then
when $h \le N$, we have that $P(X=h)=1$. When $h > N$, we have that
\bee
& P(X=N) = p^N, \nnb \\
& \resizebox{.99\hsize}{!}{$P(X=N+k) = \binom{N+k-1}{k} (1-p)^k p^N, \forall k \in [h-N-1]$} \nnb \\
& \resizebox{.99\hsize}{!}{$P(X=h) = \sum\limits_{i=0}^{N-1} \binom{h}{i} (1-p)^{h-i} p^i + \binom{h-1}{h-N} (1-p)^{h-N} p^N$}.\nnb
\eee
Then when $h \le N$, stability condition \eqref{equ:stability-condition-capacity-region-symmetric-simplify} becomes,
\be
\frac{1-q_{\max} ({A}, h)}{p}  < \frac{h}{N}.
\label{equ:stability-condition-symmetric-simplify-h-le-N}
\ee
Otherwise, when $h > N$, stability condition \eqref{equ:stability-condition-capacity-region-symmetric-simplify} becomes,
\bee
& \frac{1-q_{\max} ({A}, h)}{p}  < \frac{\mathbb{E}[X]}{N}  = \nnb \\
& \frac{NP(X\hspace{-0.6mm}=\hspace{-0.6mm}N)\hspace{-0.6mm}+\hspace{-0.6mm}\sum\limits_{k=1}^{h-N-1}(N+k)P(X=N+k) \hspace{-0.6mm}+\hspace{-0.6mm} h P(X=h)}{N} \nnb \\
& = \frac{Np^N + \sum_{k=1}^{h-N-1}(N+k)\binom{N+k-1}{k} (1-p)^k p^N}{N} \nnb \\
& + \frac{h\left(\sum\limits_{i=0}^{N-1} \binom{h}{i} (1-p)^{h-i} p^i + \binom{h-1}{h-N} (1-p)^{h-N} p^N\right)}{N}.
\label{equ:stability-condition-symmetric-simplify-h-g-N}
\eee

Note that when $q=1$, i.e., in the perfect-channel case, the stability condition \eqref{equ:stability-condition-symmetric-simplify-h-le-N} for $h \le N$ becomes
\be
N(1-q_{\max} ({A}, h)) < h,
\label{equ:stability-condition-symmetric-and-perfect-simplify-h-le-N}
\ee
which coincides with \eqref{equ:stablity-condition-perfect-channel} under the symmetric structure;
the stability condition \eqref{equ:stability-condition-symmetric-simplify-h-g-N} for $h > N$ becomes
\be
1-q_{\max} ({A}, h) < 1,
\label{equ:stability-condition-symmetric-and-perfect-simplify-h-g-N}
\ee
which always holds. This result again coincides with the analysis in Sec.~\ref{sec:perfect-channel} that
the system can always be stabilized when $h > N$.

Readers may wonder whether one can have a result similar to Theorem~\ref{the:h-min-perfect} for the non-perfect-channel case.
However, it turns out that this is not possible, as shown later in Fig.~\ref{fig:stability-symmetric-structure-h-p} in the simulation section.
This shows that the effect of the sampling period $h$  to the control system and the network system in the symmetric-structure case is much more complicated
than that in the perfect-channel case.

\section{A Monotonic Property in Terms of Channel Quality Vector} \label{sec:monotonic}
Intuitively, with better channel quality (i.e., $\bm{p}$ is larger),
packets can have more chance to be delivered successfully and thus
we can decrease the control packet dropout rate. Then for our control systems,
it becomes easier to be stabilized. We thus present the following result.

\begin{theorem} \label{thm:monotonicity-channel-quality}
Consider two channel quality vectors
$\bm{p}=(p_1,p_2,\cdots,p_N)$ and $\bm{\tilde{p}}=(\tilde{p}_1,\tilde{p}_2, \cdots, \tilde{p}_N)$
satisfying $\bm{p} \le \bm{\tilde{p}}$, i.e., $p_i \le \tilde{p}_i, \forall i \in [N]$.
Then the capacity region under channel quality vector
$\bm{p}$ is a subset of the capacity region under channel quality vector $\bm{\tilde{p}}$,
i.e., $\mathcal{R}(\bm{p},h) \subset \mathcal{R}(\bm{\tilde{p}},h)$.
\end{theorem}
\begin{IEEEproof}
With a straightforward induction procedure, instead of improving the quality of all channels in Theorem \ref{thm:monotonicity-channel-quality},
it suffices to prove it when improving the quality of only one channel (say channel 1 without loss of generality).
Therefore, we only need to prove the following result:
\begin{itemize}
\item (a) Given $\bm{p}=(p_1,p_2,\cdots,p_N)$ and $\tilde{\bm{p}}=(\tilde{p}_1,p_2,\cdots, p_N)$ where $p_1 \le \tilde{p}_1$,
we have $\mathcal{R}(\bm{p},h) \subset \mathcal{R}(\tilde{\bm{p}},h)$.
\end{itemize}

We use the idle-time-based capacity region characterization \eqref{equ:capacity-region} to prove result (a).
We reorganize \eqref{equ:capacity-region} as follows,
\be
\sum_{i \in \mathcal{S}} \frac{R_i}{p_i}  \le h - \mathbb{E}[I_{\mathcal{S}}], \quad \forall \mathcal{S} \subset [N].
\label{equ:app-capacity-region-reorganze}
\ee
Note that both LHS and RHS of \eqref{equ:app-capacity-region-reorganze} depends on the channel quality vector
$\bm{p}=(p_1,p_2,\cdots,p_N)$.
Define
\be
X_{\mathcal{S}} \triangleq h - I_{\mathcal{S}},
\label{equ:def-X-S}
\ee
which is the number of active slots in a frame if we use a work-conserving policy to scheduling
all flows in $\mathcal{S}$. According to our definition for $I_{\mathcal{S}}$ in \eqref{equ:def-I-S},
we can see that
\be
X_{\mathcal{S}} = \min\left\{\sum_{i \in \mathcal{S}} \Theta_i, h\right\}.
\ee
We next show the following lemma.

\begin{lemma} \label{lem:app-f(p)-increase}
For any $\mathcal{S} \subset \{2,3,\cdots,N\}$,
\be
f(p) = p_1 \mathbb{E}[X_{1 \cup \mathcal{S}}-X_{\mathcal{S}}],
\ee
increases as $p_1$ increases\footnote{Here for simplicity, we use $X_1$ to represent $X_{\{1\}}$, i.e., $\mathcal{S}=\{1\}$ in \eqref{equ:def-X-S}.
Similarly, we use $X_{1 \cup \mathcal{S}}$ to represent $X_{\{1\} \cup \mathcal{S}}$.}.
\end{lemma}
The proof of Lemma~\ref{lem:app-f(p)-increase} is shown in Appendix~\ref{app:proof-of-app-f(p)-increase}.

According to \eqref{equ:app-capacity-region-reorganze}, the capacity region  under channel quality vector $\bm{p}=(p_1,p_2,\cdots,p_N)$, i.e., $\mathcal{R}(\bm{p},h)$ becomes
\be
\sum_{i \in \mathcal{S}} \frac{R_i}{p_i}  \le \mathbb{E}[X_{\mathcal{S}}], \quad \forall \mathcal{S} \subset [N].
\label{equ:app-capacity-region-p-1}
\ee
Depending on whether $\mathcal{S}$ in \eqref{equ:app-capacity-region-p-1} contains flow 1 or not, we can equivalently express \eqref{equ:app-capacity-region-p-1} as
\bse
\bee
& \sum_{i \in \mathcal{S}} \frac{R_i}{p_i}  \le \mathbb{E}[{X}_{\mathcal{S}}], \quad \forall \mathcal{S} \subset \{2,3,\cdots,N\},  \label{equ:app-capacity-region-p-2-1} \\
&  \frac{R_1}{{p}_1} + \sum_{i \in \mathcal{S}} \frac{R_i}{p_i}  \le \mathbb{E}[{X}_{1 \cup \mathcal{S}}], \quad \forall \mathcal{S} \subset \{2,3,\cdots,N\}, \label{equ:app-capacity-region-p-2-2}
\eee
\label{equ:app-capacity-region-p-2}
\ese

Under the improved channel quality  vector $\bm{\tilde{p}}=(\tilde{p}_1,p_2,\cdots,p_N)$, we use
$\tilde{X}_{\mathcal{S}}$ to denote the number of active slots in a frame if we use a work-conserving policy to scheduling
all flows in $\mathcal{S}$. Then the new capacity region $\mathcal{R}(\tilde{\bm{p}},h)$ becomes,
\bse
\bee
& \sum_{i \in \mathcal{S}} \frac{R_i}{p_i}  \le \mathbb{E}[\tilde{X}_{\mathcal{S}}] = \mathbb{E}[{X}_{\mathcal{S}}],  \quad \forall \mathcal{S} \subset \{2,3,\cdots,N\}\label{equ:app-capacity-region-p-tilde-1} \\
&  \frac{R_1}{\tilde{p}_1} + \sum_{i \in \mathcal{S}} \frac{R_i}{p_i}  \le \mathbb{E}[\tilde{X}_{1 \cup \mathcal{S}}], \quad \forall \mathcal{S} \subset \{2,3,\cdots,N\}, \label{equ:app-capacity-region-p-tilde-2}
\eee
\label{equ:app-capacity-region-p-tilde}
\ese

We now show that any $\bm{R}=(R_1,R_2,\cdots,R_N)$ satisfying
\eqref{equ:app-capacity-region-p-2} also satisfies \eqref{equ:app-capacity-region-p-tilde}.
Since $\bm{R}$ satisfies \eqref{equ:app-capacity-region-p-2-1}, $\bm{R}$ also satisfies
\eqref{equ:app-capacity-region-p-tilde-1}.
According to Lemma~\ref{lem:app-f(p)-increase}, since $p_1 \le \tilde{p}_1$, we have
\be
 p_1 \mathbb{E}[X_{1 \cup \mathcal{S}}-X_{\mathcal{S}}] \le  \tilde{p}_1 \mathbb{E}[\tilde{X}_{1 \cup \mathcal{S}}- \tilde{X}_{\mathcal{S}}],
\ee
Since from \eqref{equ:app-capacity-region-p-2-1} and \eqref{equ:app-capacity-region-p-tilde-1} we have
\be
\sum_{i \in \mathcal{S}} \frac{R_i}{p_i} \le \mathbb{E}[\tilde{X}_{\mathcal{S}}] = \mathbb{E}[{X}_{\mathcal{S}}],
\ee
then
\bee
&  p_1 \left(\mathbb{E} [X_{1 \cup \mathcal{S}}] - \sum_{i \in \mathcal{S}} \frac{R_i}{p_i}\right) -  \tilde{p}_1 \left( \mathbb{E}[\tilde{X}_{1 \cup \mathcal{S}}] - \sum_{i \in \mathcal{S}} \frac{R_i}{p_i}\right) \nnb \\
& = p_1 \mathbb{E} [X_{1 \cup \mathcal{S}}] - \tilde{p}_1 \mathbb{E} [\tilde{X}_{1 \cup \mathcal{S}}] + (\tilde{p_1} - p_1)  \sum_{i \in \mathcal{S}} \frac{R_i}{p_i} \nnb \\
& \le p_1 \mathbb{E} [X_{1 \cup \mathcal{S}}] - \tilde{p}_1 \mathbb{E} [\tilde{X}_{1 \cup \mathcal{S}}] + (\tilde{p_1} - p_1)  \mathbb{E} [{X}_{\mathcal{S}}] \nnb \\
& =  p_1 \mathbb{E}[X_{1 \cup \mathcal{S}}-X_{\mathcal{S}}]  - \tilde{p}_1 \mathbb{E}[\tilde{X}_{1 \cup \mathcal{S}}- \tilde{X}_{\mathcal{S}}] \nnb \\
& \le 0.
\eee
Therefore, we have
\be
 p_1 \left(\mathbb{E} [X_{1 \cup \mathcal{S}}] - \sum_{i \in \mathcal{S}} \frac{R_i}{p_i}\right) \le  \tilde{p}_1 \left( \mathbb{E}[\tilde{X}_{1 \cup \mathcal{S}}] - \sum_{i \in \mathcal{S}} \frac{R_i}{p_i}\right).
\label{equ:app-key-inequ-1}
\ee
Note that \eqref{equ:app-capacity-region-p-2-2} can be reorganized as
\be
R_1 \le  p_1 \left(\mathbb{E} [X_{1 \cup \mathcal{S}}] - \sum_{i \in \mathcal{S}} \frac{R_i}{p_i}\right).
\ee
Due to \eqref{equ:app-key-inequ-1}, we thus have
\be
R_1 \le  \tilde{p}_1 \left( \mathbb{E}[\tilde{X}_{1 \cup \mathcal{S}}] - \sum_{i \in \mathcal{S}} \frac{R_i}{p_i}\right) ,
\ee
which is equivalent to \eqref{equ:app-capacity-region-p-tilde-2}. Therefore, \eqref{equ:app-capacity-region-p-tilde-2} also holds and thus \eqref{equ:app-capacity-region-p-tilde}
holds.

Therefore, any $\bm{R}=(R_1,R_2,\cdots,R_N)$ satisfying
\eqref{equ:app-capacity-region-p-2} also satisfies \eqref{equ:app-capacity-region-p-tilde}, implying $\mathcal{R}(\bm{p},h) \subset \mathcal{R}(\bm{\tilde{p}},h)$.

This completes the proof.
\end{IEEEproof}

Theorem~\ref{thm:general-stabiliyt-condition} and Theorem~\ref{thm:monotonicity-channel-quality} show that
if the WNCS can be stabilized under channel quality vector $\bm{p}$, it can also be stabilized
under a better channel quality vector $\bm{\tilde{p}}$ where $\bm{p} \le \bm{\tilde{p}}$.
As a by-product, in the symmetric-structure case, we can use a binary-search scheme to find
the minimum channel quality $p$ such that the WNCS can be stabilized.
In addition, we remark that Theorem~\ref{thm:monotonicity-channel-quality} itself is
a new result for the delay-constrained wireless communication problem with frame-synchronized
traffic pattern \cite{hou2009qos,hou2012queueing}.

\section{Heterogenous Sampling Periods} \label{sec:heterogenous-sampling-periods}
\begin{figure}
  \centering
  \includegraphics[width=\linewidth]{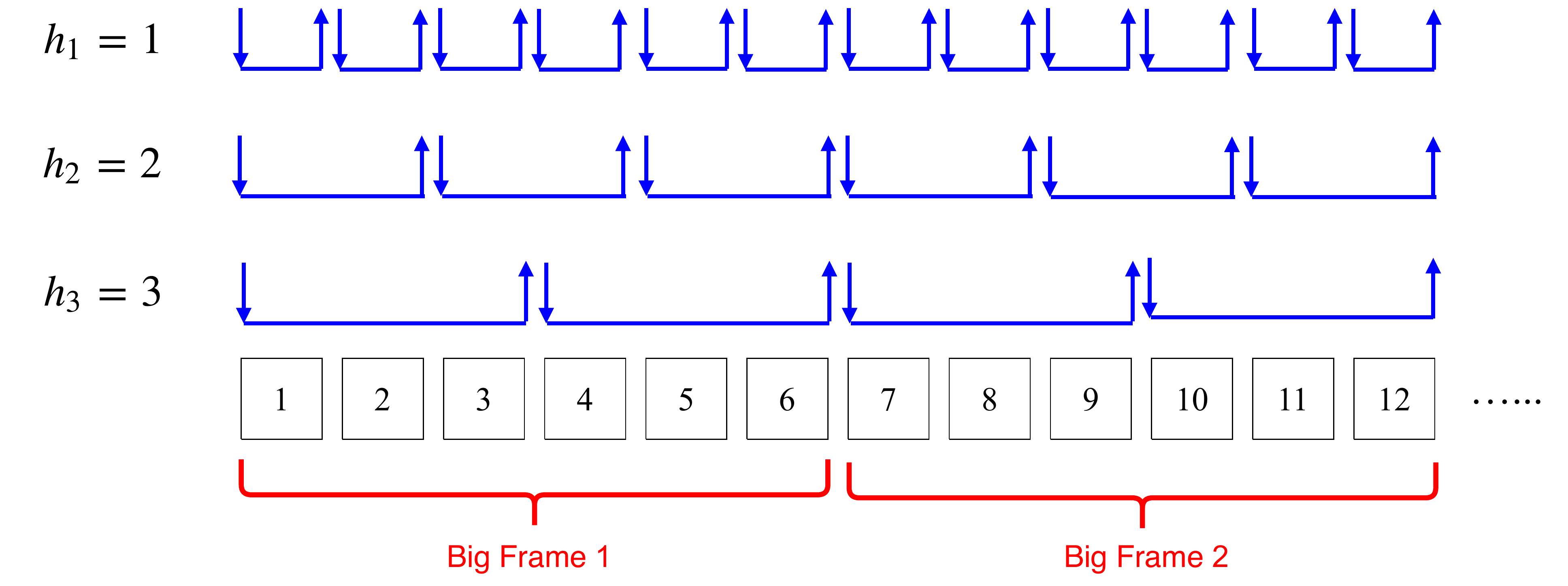}\\
  \caption{An example for heterogenous sampling periods.}\label{fig:traffic-multiple-h}
\end{figure}

In our model, we assume that all sub-systems are sampled every $h$ slots. It is
more practical to consider heterogenous sampling periods. Suppose that sub-system
$i$ is sampled every $h_i \in \mathbb{Z}^+$ slots. An example of $N=3, h_1=1, h_2=2, h_3=3$ is shown
in Fig.~\ref{fig:traffic-multiple-h}.
Again, our aim is to study the
stability condition of the whole system.

Since the traffic pattern is no longer frame-synchronized,
we cannot use the idle-time-based capacity region characterization  \cite{hou2009qos,hou2012queueing}.
However, the MDP-based approach \cite{deng2017timely} can be applied to general traffic patterns including
our case with heterogenous sampling periods. In particular, consider the  common period
\be
H = \textsf{Least Common Multiple}(h_1,h_2,\cdots,h_N).
\ee
Then every $H$ slots forms a \emph{big frame}. Namely, big frame 1 is from slot 1
to slot $H$; big frame 2 is from slot $H+1$ to slot $2H$; and so on (see Fig.~\ref{fig:traffic-multiple-h}).
Clearly, in a big frame, sub-system $i$ will be sampled $H/h_i$ times. Thus,
every big frame has $n_i=H/h_i$ periods of sub-system $i$. Then we can characterize the timely
capacity region $\mathcal{R}(\bm{p}, \bm{h})$ by some linear equalities/inequalities (see \cite[Equ.~(11)]{deng2017timely}), similar to \eqref{equ:capacit-region-MDP} for frame-synchronized traffic pattern.
In addition, for any achievable timely throughput vector $\bm{R} \in \mathcal{R}(\bm{p}, \bm{h})$, we can design a \emph{cyclo-periodic}
scheduling policy to achieve it. Since we consider heterogenous sampling periods, the constructed
cyclo-periodic scheduling policy cannot ensure that $\{\gamma^i_k:k=1,2,\cdots\}$ are i.i.d. Instead,
the constructed cyclo-periodic scheduling policy satisfies the following conditions,
\begin{itemize}
\item (a) The random variables $\{\gamma^i_{j}: j=1,2,\cdots\}$ are independent (but not identical) distributed.
\item (b) For any $i \in [N]$ and any $j \in [n_i]$, the random variables $\{\gamma^i_{(k-1)n_i+j}: k=1,2, \cdots \}$ are i.i.d.
i.e., the delivery probability of the control message is the same every $n_i$ periods, i.e.,
\be
P(\gamma^i_{(k-1)n_i+j}=1) = \delta^i_j, \quad \forall k=1,2,\cdots.
\ee
\end{itemize}

\begin{figure}
  \centering
  \includegraphics[width=\linewidth]{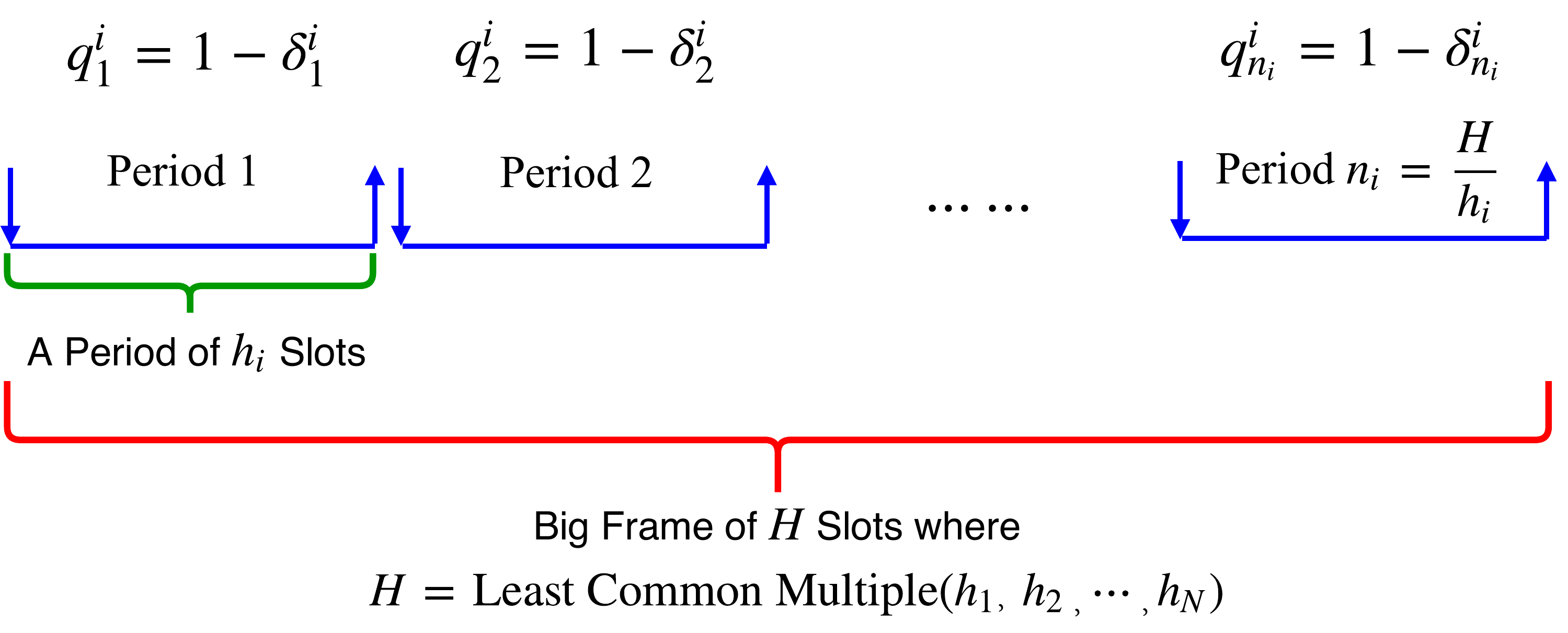}\\
  \caption{Illustration of the notations in heterogenous sampling periods.}\label{fig:traffic-multiple-h-notation}
\end{figure}

Therefore, random variables $\{\gamma^i_{j}: j=1,2,\cdots\}$ are no longer i.i.d now.
We call that $\{\gamma^i_{j}: q=1,2,\cdots\}$ are \emph{independent and periodic distributed} (i.p.d) with period $n_i$
if they satisfy (a) and (b) above.
We also denote $q^i_j = P(\gamma^i_{(k-1) n_i+j}=0)=1-\delta^i_j$, which is
the dropout rate of sub-system $i$ in the $j$-th period (and any $((k-1) n_i+j)$-th period).
 We illustrate the notations introduced above in
Fig.~\ref{fig:traffic-multiple-h-notation}.
We now characterize a sufficient condition to stabilize sub-system $i$ under i.p.d. dropout out events.

\begin{theorem} \label{thm:stability-ipd}
Suppose that random variables $\{\gamma^i_{j}: j=1,2,\cdots\}$ are i.p.d. with period $n_i$.
Then sub-system $i$ is (mean-square) stable if $q^i_{j}<q^i_{\max}({A}_i,h_i), \forall j \in [n_i],$
where $q^i_{\max}({A}_i,h_i)$ is defined in \eqref{equ:max-dropout-rate}.
\end{theorem}
\begin{IEEEproof}
Please see Appendix~\ref{app:proof-of-stability-ipd}.
\end{IEEEproof}

\begin{figure*}[t]
\hspace{-0.5cm}
  \begin{minipage}[t]{0.01\linewidth}~~~
  \end{minipage}
  \begin{minipage}[t]{0.32\linewidth}
    \centering
\includegraphics[width = \linewidth]{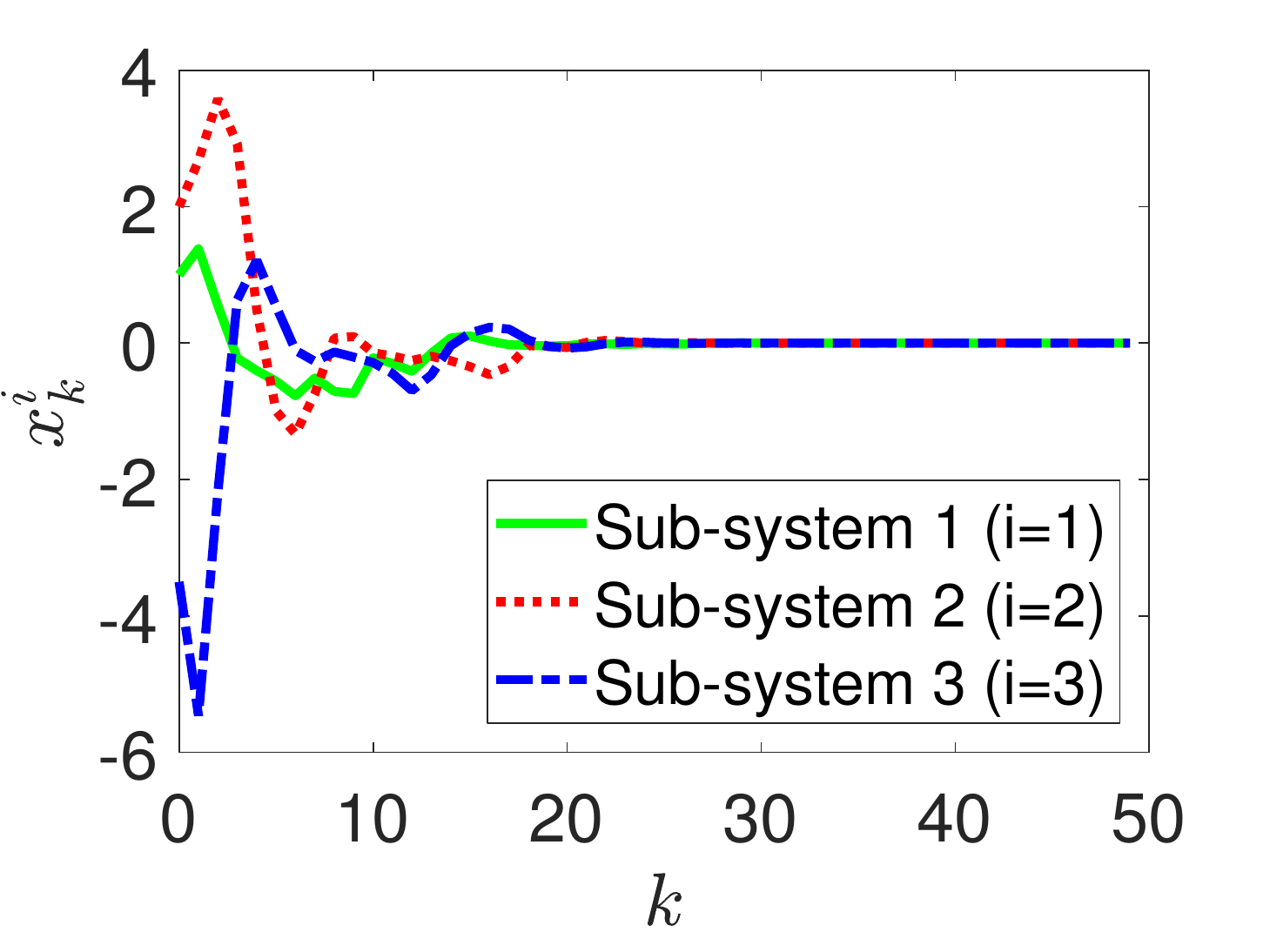}\\
   \caption{State evolution of three sub-systems with system parameters in \eqref{equ:sim-parameter1}. \label{fig:state-evolution}}
  \label{fig:rounded-fuel-curve.}
  \end{minipage}
  \begin{minipage}[t]{0.01\linewidth}~~~
  \end{minipage}
  \begin{minipage}[t]{0.32\linewidth}
    \centering
    \includegraphics[width = \linewidth]{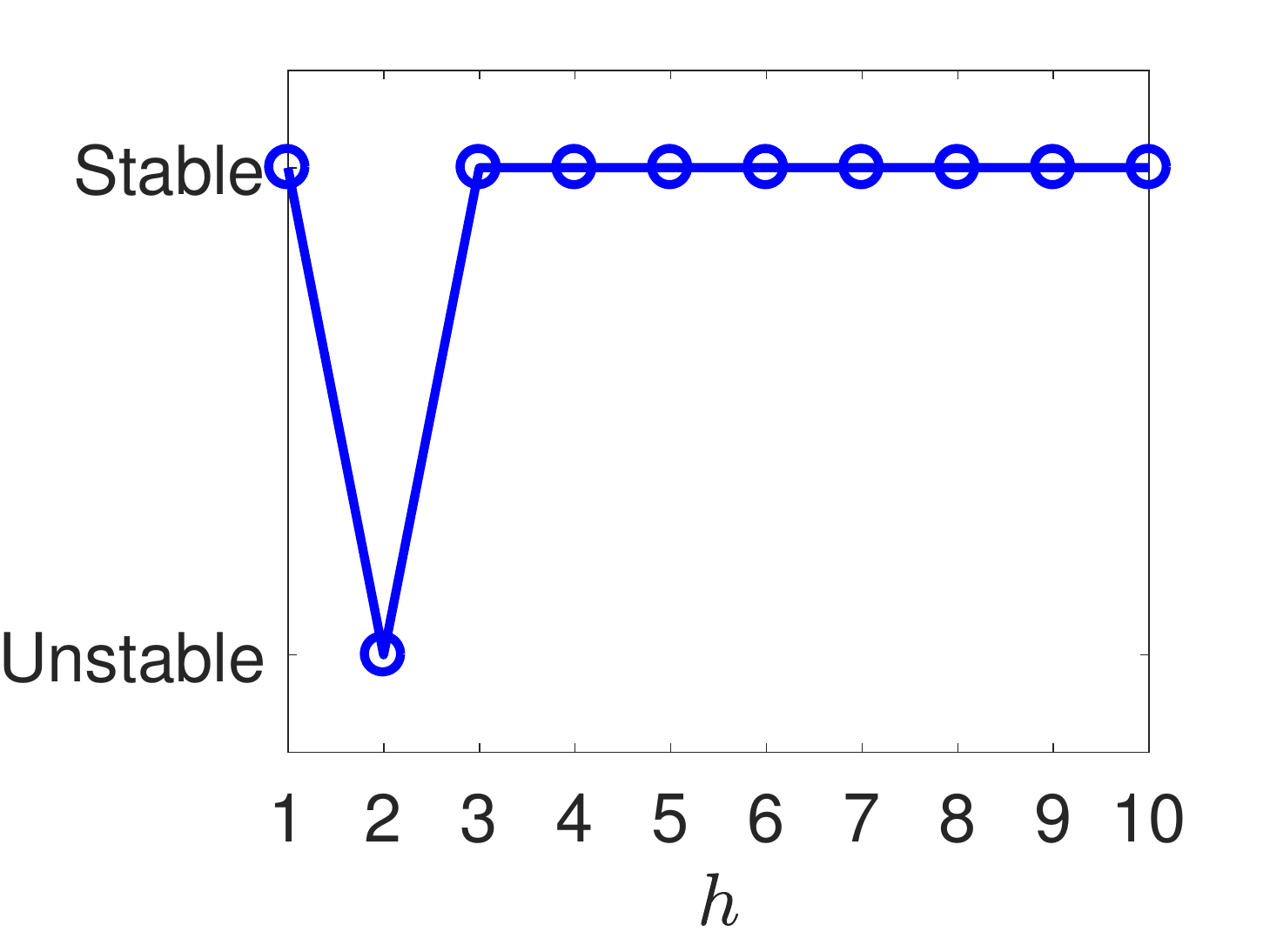}\\
   \caption{Stability condition for perfect-channel case  with system parameters in \eqref{equ:sim-parameter2}.  \label{fig:stability-perfect-channel-h}}
\end{minipage}
  \begin{minipage}[t]{0.01\linewidth}~~~
  \end{minipage}
\begin{minipage}[t]{0.32\linewidth}
    \centering
  \includegraphics[width=\linewidth]{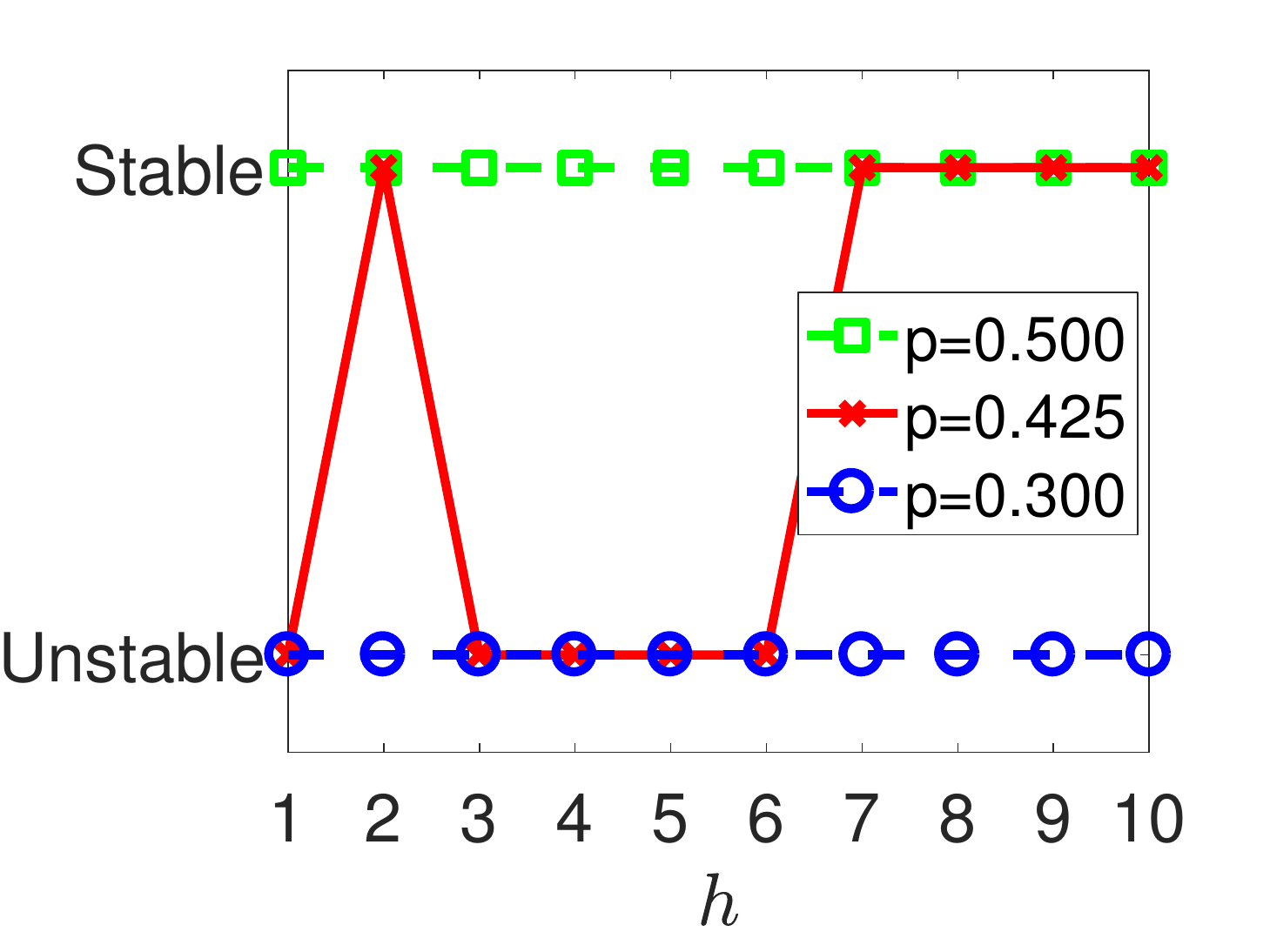}\\
  \caption{Stability condition for symmetric-structure case  with system parameters in \eqref{equ:sim-parameter3}.
   \label{fig:stability-symmetric-structure-h-p}}
  \end{minipage}
  \begin{minipage}[t]{0.01\linewidth}~~~
  \end{minipage}
\end{figure*}

Theorem~\ref{thm:stability-ipd} requires that the dropout rate of any period is strictly less than
$q^i_{\max}({A}_i,h_i)$. This adds further restrictions to the design of the network scheduling policy.
To further propose a sufficient condition to stabilize the whole system, we need to utilize and modify the details of
the MDP-based capacity region characterization in \cite{deng2017timely}.
\begin{theorem} \label{thm:sufficient-condition-for-heterogenous-sampling-period}
The system with heterogenous sampling periods $h_1,h_2, \cdots, h_N$ can be stabilized if
there exists a set of $\{x_{t}(\bm{s},a)\}$  such that the following linear inequalities hold,
\bse \label{equ:capacit-region-MDP-Heterogenous}
\bee
& \sum_{a \in \mathcal{A}} x_{t+1}(\bm{s}',a) = \sum_{\bm{s} \in \mathcal{S}} \sum_{a \in \mathcal{A}} x_t(\bm{s},a) P_t(\bm{s}' | \bm{s}, a), \nnb \\
& \qquad \qquad \forall \bm{s}' \in \mathcal{S}, t \in [H-1], \\
& \sum_{a \in \mathcal{A}} x_{1}(\bm{s}',a) = \sum_{\bm{s} \in \mathcal{S}} \sum_{a \in \mathcal{A}} x_H(\bm{s},a) P_H(\bm{s}' | \bm{s}, a), \nnb \\
& \qquad \qquad \forall \bm{s}' \in \mathcal{S}, \\
& 1- q^i_{\max}({A}_i,h_i) < \sum_{t=(j-1)h_i+1}^{jh_i} \sum_{\bm{s} \in \mathcal{S}} \sum_{a \in \mathcal{A}} {x_t(\bm{s},a)r_i(\bm{s},a)}, \nnb \\
& \qquad \qquad \forall i \in [N], j \in [n_i],  \label{equ:capacit-region-MDP-Heterogenous-con3}\\
&  \sum_{\bm{s} \in \mathcal{S}} \sum_{a \in \mathcal{A}} x_t(\bm{s},a) =1, \quad \forall t \in [H], \\
&  x_t(\bm{s},a) \ge 0,  \quad \forall t \in [H], \bm{s} \in \mathcal{S}, a \in \mathcal{A}.
\eee
\ese
\end{theorem}
\begin{IEEEproof}
If \eqref{equ:capacit-region-MDP-Heterogenous} holds, we obtain a set of $\{x_t(\bm{s},a)\}$.
Then we construct the cyclo-periodic  policy according to \cite[Equ.~(12)]{deng2017timely}, similar to \eqref{equ:RAC-scheduling-policy} for frame-synchronized traffic pattern.
It is straightforward to show that $\{\gamma^i_j:j=1,2,\cdots\}$ are independent because the state of any sub-system $i$ starts over with state $S_i=1$ at the beginning of each period.
In addition, since the state of the whole system starts over with $\bm{s}=(1,1,\cdots,1)$ at the beginning of each big frame and the scheduling policy is cyclo-periodic,
the induced $\{\gamma^i_{(k-1) n_i+j}: k=1,2, \cdots \}$ are i.i.d. for any $i \in [N]$ and any $j \in [n_i]$.
Thus, random variables $\{\gamma^i_j:j=1,2,\cdots\}$ are i.p.d. with period $n_i$.
In addition, due to \eqref{equ:capacit-region-MDP-Heterogenous-con3}, the dropout rate of the sub-system $i$'s control message in any period $j \in [n_i]$ satisfies
\be
q_j^i = 1 - \sum_{t=(j-1)h_i+1}^{jh_i} \sum_{\bm{s} \in \mathcal{S}} \sum_{a \in \mathcal{A}} {x_t(\bm{s},a)r_i(\bm{s},a)} < q^i_{\max}({A}_i,h_i). \nnb
\ee
Therefore, according to Theorem~\ref{thm:stability-ipd}, the whole system is (mean-square) stable.
\end{IEEEproof}

Theorem \ref{thm:sufficient-condition-for-heterogenous-sampling-period} characterizes a sufficient condition for stability in the case of heterogenous sampling periods.
But we should note that it is not a necessary condition. How to find the exact stability condition in the case of heterogenous sampling periods
is an interesting and promising future direction.

\section{Simulation} \label{sec:simulation}
In this section, we use simulation to confirm our theoretic analysis.

\textbf{General System.}
We consider a general (imperfect-channel asymmetric-structure) system with system parameters,
\bee
& N=3, h=5, \Delta=0.01, \nnb \\
& A_1=6.5137, A_2= 5.8265, A_3=8.8964, \nnb \\
& B_1=1, B_2=1, B_3=1, \nnb \\
& p_1=0.7690, p_2=0.7277, p_3=0.2846.
\label{equ:sim-parameter1}
\eee
According to our stability condition \eqref{equ:stability-condition-capacity-region}, we can check that
the system can be stabilized. We then construct the network scheduling policy and the control policy
to get the per-frame state of each sub-system, which is shown in Fig.~\ref{fig:state-evolution}. As we can see, indeed, the states of all three
sub-systems converge to 0 and thus all three sub-systems are stabilized. This verifies our stability condition \eqref{equ:stability-condition-capacity-region}
for general system.

\textbf{Perfect-Channel Case.}
We further consider a perfect-channel case with system parameters,
\bee
& N=6, \Delta= 0.0114, \nnb \\
& \bm{A}=(A_1,A_2,A_3,A_4,A_5,A_6) \nnb \\
&  \;\;\; =(3.7482, 8.7512, 7.7711, 8.5482, 6.8823, 5.6830), \nnb \\
& \bm{p}=(p_1, p_2, p_3, p_4, p_5, p_6)=(1,1,1,1,1,1).
\label{equ:sim-parameter2}
\eee
We change the sampling period $h$ from 1 slot to 10 slots. The stability result
is shown in Fig.~\ref{fig:stability-perfect-channel-h}. We can see that there exists an $h_{\min}=3 \in \{1,2,3,4,5,6\}$
such that the system can be stabilized if $h \ge h_{\min}$. This is consistent with Theorem~\ref{the:h-min-perfect}. From this figure, we can
also see that $h \ge h_{\min}$ is not necessary for stability in \eqref{equ:stablity-condition-perfect-channel}, because
the system can be stabilized when $h=1 < h_{\min}=3$.

\textbf{Symmetric-Structure Case.}
We next consider a symmetric-structure (imperfect-channel) case with parameters,
\bee
N=2, \Delta=0.1, A_1=A_2=A=1.
\label{equ:sim-parameter3}
\eee
We change the sampling period $h$ from 1 slot to 10 slots and consider three different levels of channel quality $p_1=p_2=p \in \{0.300, 0.425, 0.500\}$.
The stability result is shown in Fig.~\ref{fig:stability-symmetric-structure-h-p}. We can see that the system is unstable for all sampling periods
when the channel quality is bad, i.e., $p=0.30$. However, the system can be stabilized for all sampling periods when the channel quality is
good, i.e., $p=0.5$. When the channel quality is medium, i.e., $p=0.425$, the system can be stabilized when the sampling period $h \in \{2,7,8,9,10\}$
and unstable otherwise. Thus, for such imperfect-channel case, we do not have a similar result like Theorem~\ref{the:h-min-perfect} for perfect-channel case.
Instead, the stability result becomes more complicated when we consider the effect of channel quality.

\begin{figure*}[t]
\hspace{-0.5cm}
  \begin{minipage}[t]{0.01\linewidth}~~~
  \end{minipage}
 \begin{minipage}[t]{0.32\linewidth}
    \centering
   \includegraphics[width=\linewidth]{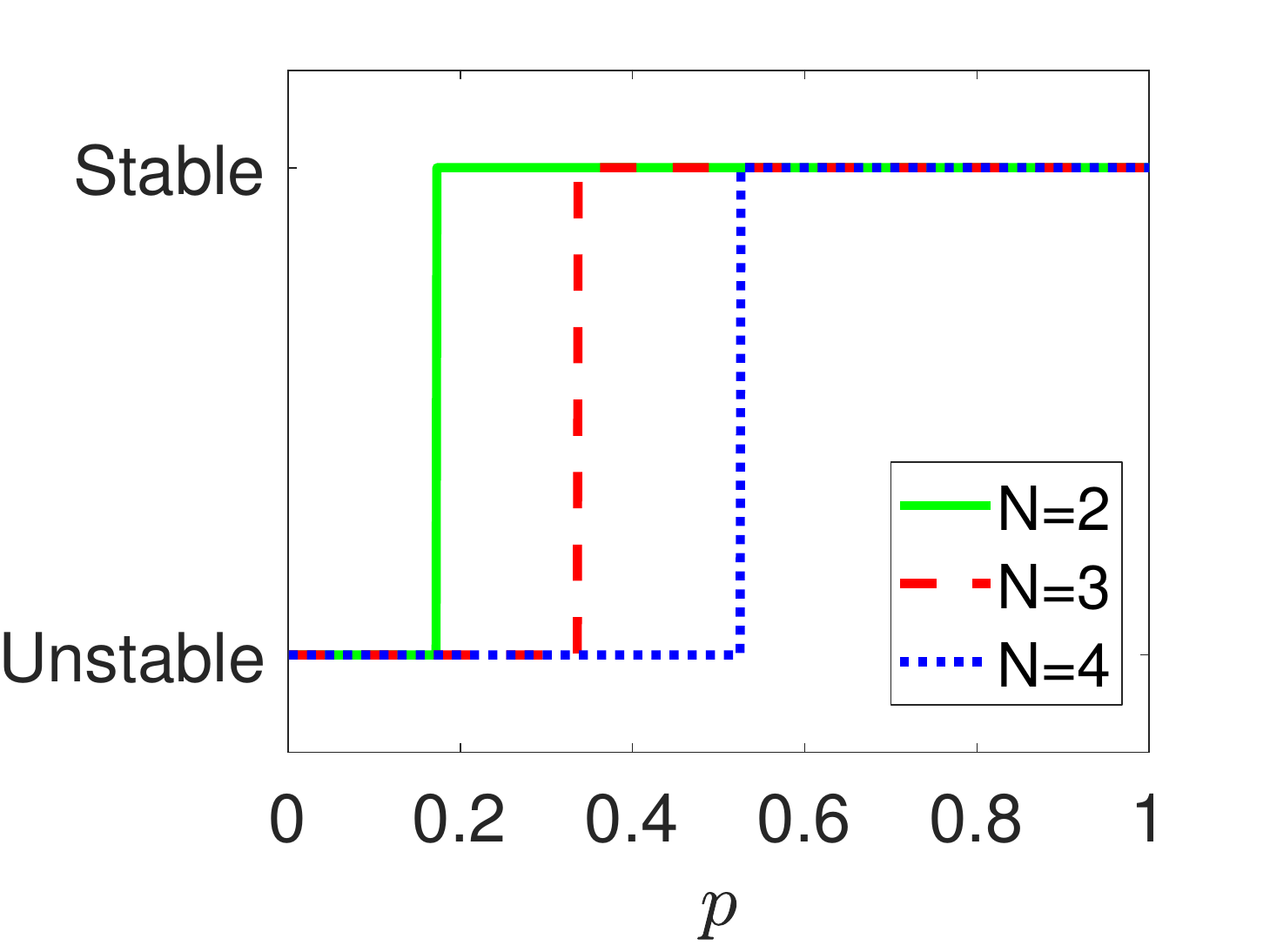}\\
  \caption{The monotonic property in terms of channel quality  with system parameters in \eqref{equ:sim-parameter4-1}-\eqref{equ:sim-parameter4-3}. \label{fig:stability-monotonic-N-p}}
  \end{minipage}
  \begin{minipage}[t]{0.01\linewidth}~~~
  \end{minipage}
  \begin{minipage}[t]{0.32\linewidth}
    \centering
  \includegraphics[width=\linewidth]{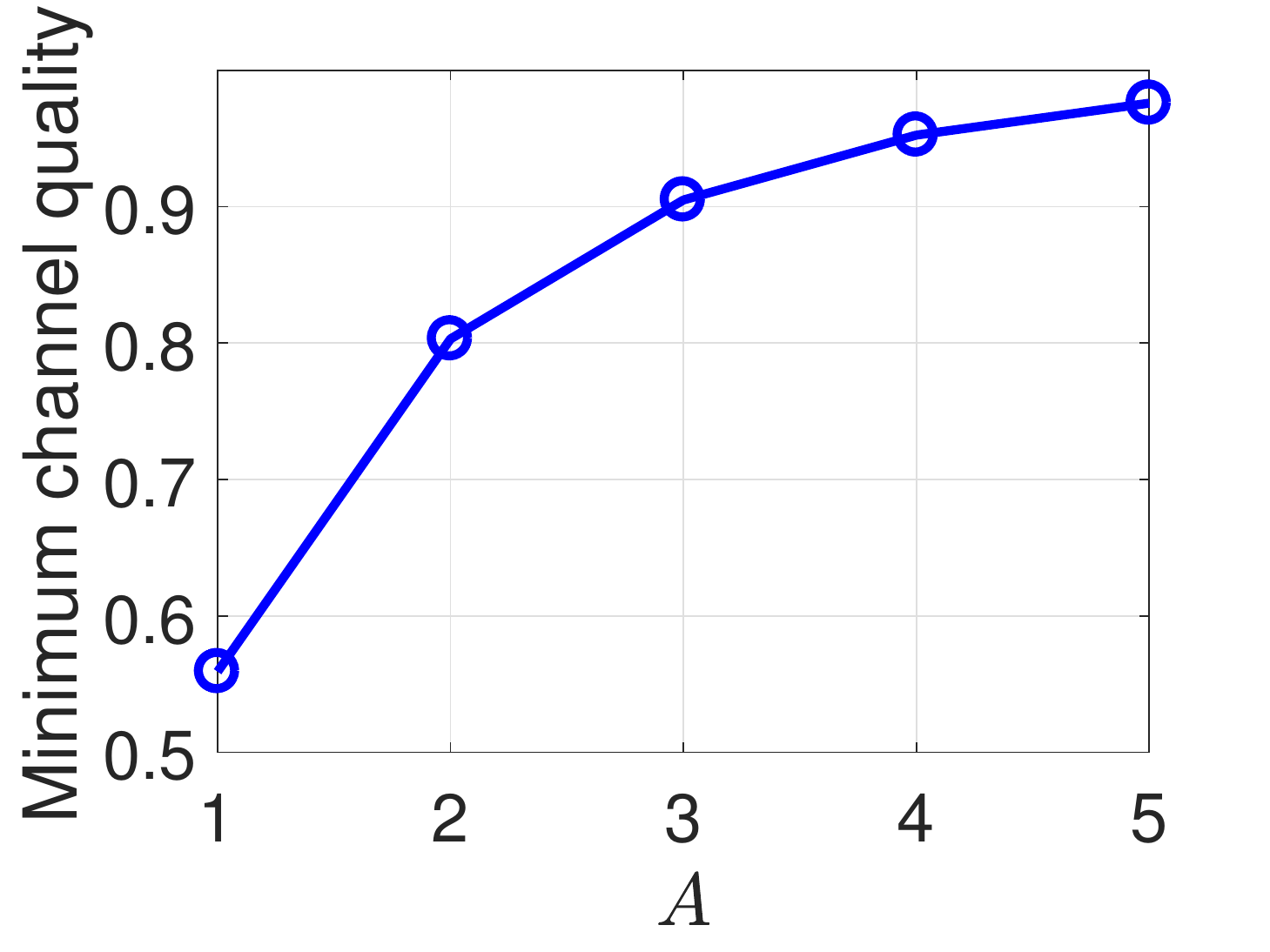}\\
  \caption{The minimal channel quality  to stabilize the system (v.s. $A$) with system parameters  in \eqref{equ:sim-parameter5}.
   \label{fig:min-p-A}}
  \end{minipage}
  \begin{minipage}[t]{0.01\linewidth}~~~
  \end{minipage}
  \begin{minipage}[t]{0.32\linewidth}
    \centering
   \includegraphics[width=\linewidth]{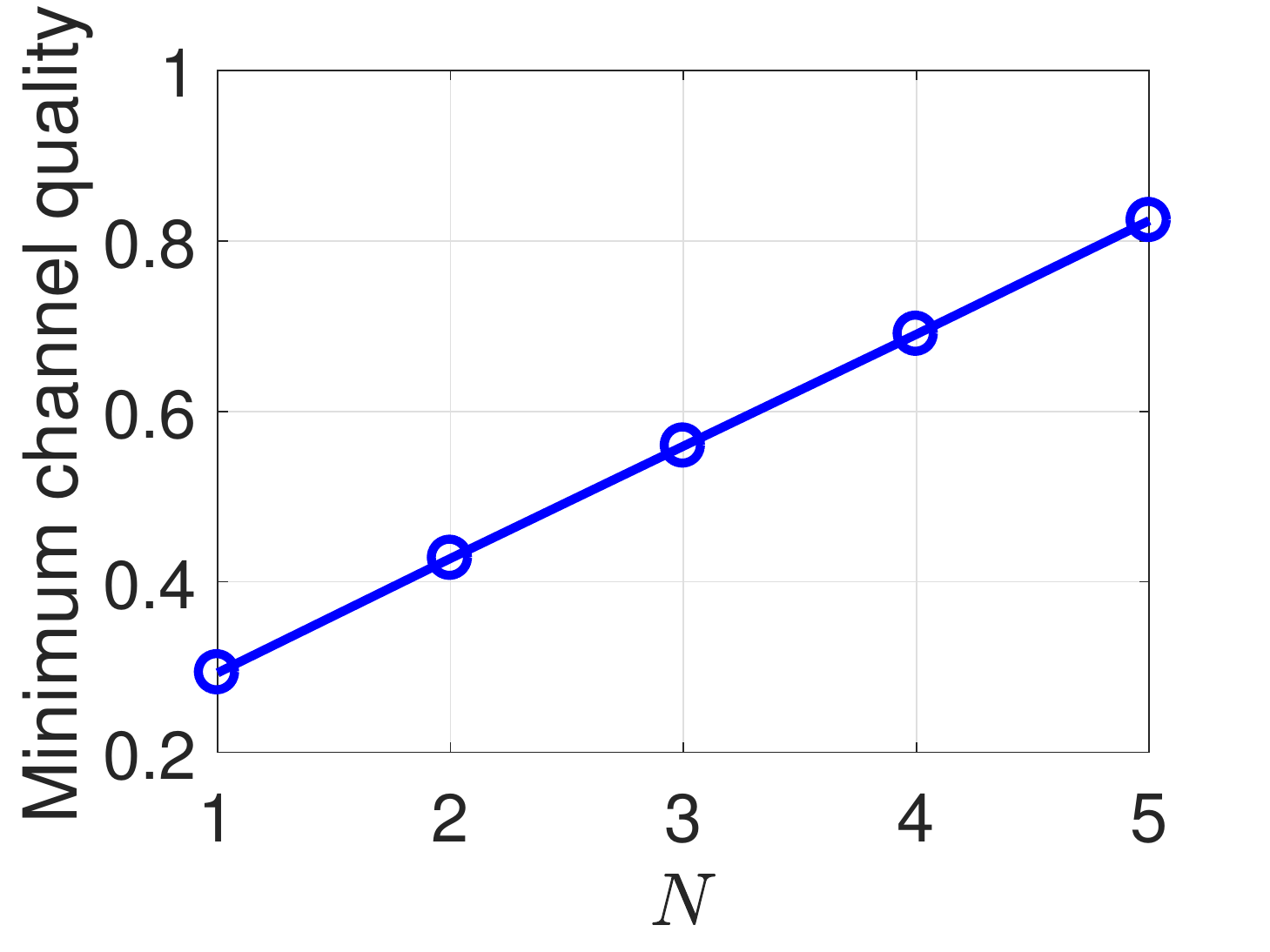}\\
  \caption{The minimal channel quality  to stabilize the system  (v.s. $N$) with system parameters in \eqref{equ:sim-parameter6}. \label{fig:min-p-N}}
  \end{minipage}
  \begin{minipage}[t]{0.01\linewidth}~~~
  \end{minipage}
\end{figure*}

\textbf{Monotonic Result.}
We verify our monotonic result (Theorem~\ref{thm:monotonicity-channel-quality}) by considering the following three systems with different $N$'s,
\begin{itemize}
\item System 1 with $N=2$:
\bee
& N=2,  h=3, \Delta=0.01, \nnb \\
& \bm{A}=(A_1,A_2)=(1.9047, 6.1553), \nnb \\
& p_1=p_2=p.
\label{equ:sim-parameter4-1}
\eee
\item System 2 with $N=3$:
\bee
& N=3,  h=3, \Delta=0.01, \nnb \\
& \bm{A}=(A_1,A_2,A_3)=(1.9047, 6.1553, 7.9464), \nnb \\
& p_1=p_2=p_3=p.
\label{equ:sim-parameter4-2}
\eee
\item System 3 with $N=4$:
\bee
& N=4,  h=3, \Delta=0.01, \nnb \\
& \bm{A}=(A_1,A_2,A_3,A_4)=(1.9047, 6.1553, 7.9464, 9.3456), \nnb \\
& p_1=p_2=p_3=p_4=p.
\label{equ:sim-parameter4-3}
\eee
\end{itemize}
We change $p$ from 0.001 to 1. The stability result is shown in Fig.~\ref{fig:stability-monotonic-N-p}. As we can see, for each system,
we have a monotonic property in terms of channel quality $p$. This verifies Theorem~\ref{thm:monotonicity-channel-quality}.

\textbf{Minimum Channel Quality.}
Our monotonic result, i.e., Theorem~\ref{thm:monotonicity-channel-quality}, further enables us to find the minimum channel quality
to stabilize the whole system for the symmetric-structure case. Namely, we can use a binary-search scheme to efficiently find the minimum channel quality
\be
\resizebox{.99\linewidth}{!}{$p_{\min} \triangleq \inf\{p: \text{system can be stabilized under channel quality $p$}\}$}.
\ee

We evaluate the effect of parameter $A_1=A_2=\cdots=A_N=A$
and the effect of $N$, i.e., the number of sub-systems.
First, we consider the following system parameters in the symmetric-structure case:
\be
N=3,\Delta=0.01,h=5. \label{equ:sim-parameter5}
\ee
We change $A$ from 1 to 5 and the minimum channel quality is shown in Fig.~\ref{fig:min-p-A}.
We can see that when $A$ increases, the channel quality also needs to be improved to stabilize the system.
This is because control-system parameter $A$ (see \eqref{sys001}) represents the
degree of self-amplification and larger $A$ means larger self-instability.
Second, we consider the following system parameters in the symmetric-structure case:
\be
A=1,\Delta=0.01,h=5. \label{equ:sim-parameter6}
\ee
We change $N$ from 1 to 5 and the minimum channel quality
is shown in Fig.~\ref{fig:min-p-N}. As we cas see, the channel quality needs to be improved to stabilize the system
when the number of sub-systems increases.
This is because when we have more sub-systems to be stabilized,
virtually we need to allocate more communication resources to the new sub-systems;
thus, we need to improve the cannel quality to create more ``communication resources".

\section{Conclusion} \label{sec:conclusion}
In this paper, we characterize the stability condition of
a WNCS with multiple plants and controllers sharing a common wireless channel
under the joint design of control policy and network scheduling policy.
To solve our WNCS problem, we have leveraged the recent results in the research area of
delay-constrained wireless communication \cite{hou2009qos, hou2012queueing, deng2017timely}.
In the future, it is interesting and important to generalize our system in several aspects.
First, we have only considered the scalar-state case and thus
the general vector-state case is worth studying. Second, we assume that
the hard deadline of each control message is equal to the sampling period corresponding to
$d=1$ in \cite{cheng2017integrated}. It would also be interesting to investigate the case of $d > 1$.
Finally, our results are based on the assumption that the dropout random variables $\{\gamma_k^i:k=1,2,\cdots\}$ are i.i.d. or i.p.d.,
which simplifies the design of the control policy but shrinks the design space.
It would be challenging to consider the full design space where $\{\gamma_k^i:k=1,2,\cdots\}$ may not be i.i.d or i.p.d.

\bibliographystyle{IEEEtran}
\bibliography{ref}

\begin{thebibliography}{10}
\providecommand{\url}[1]{#1}
\csname url@samestyle\endcsname
\providecommand{\newblock}{\relax}
\providecommand{\bibinfo}[2]{#2}
\providecommand{\BIBentrySTDinterwordspacing}{\spaceskip=0pt\relax}
\providecommand{\BIBentryALTinterwordstretchfactor}{4}
\providecommand{\BIBentryALTinterwordspacing}{\spaceskip=\fontdimen2\font plus
\BIBentryALTinterwordstretchfactor\fontdimen3\font minus
  \fontdimen4\font\relax}
\providecommand{\BIBforeignlanguage}[2]{{%
\expandafter\ifx\csname l@#1\endcsname\relax
\typeout{** WARNING: IEEEtran.bst: No hyphenation pattern has been}%
\typeout{** loaded for the language `#1'. Using the pattern for}%
\typeout{** the default language instead.}%
\else
\language=\csname l@#1\endcsname
\fi
#2}}
\providecommand{\BIBdecl}{\relax}
\BIBdecl

\bibitem{deng2018stability}
L.~Deng, C.~Tan, and W.~S. Wong, ``On stability condition of wireless networked
  control systems under joint design of control policy and network scheduling
  policy,'' in \emph{Proc. IEEE CDC}, 2018.

\bibitem{gupta2010networked}
R.~A. Gupta and M.-Y. Chow, ``Networked control system: Overview and research
  trends,'' \emph{IEEE Transactions on Industrial Electronics}, vol.~57, no.~7,
  pp. 2527--2535, 2010.

\bibitem{baillieul2007control}
J.~Baillieul and P.~J. Antsaklis, ``Control and communication challenges in
  networked real-time systems,'' \emph{Proceedings of the IEEE}, vol.~95,
  no.~1, pp. 9--28, 2007.

\bibitem{hespanha2007survey}
J.~P. Hespanha, P.~Naghshtabrizi, and Y.~Xu, ``A survey of recent results in
  networked control systems,'' \emph{Proceedings of the IEEE}, vol.~95, no.~1,
  pp. 138--162, 2007.

\bibitem{al2016design}
A.~W. Al-Dabbagh and T.~Chen, ``Design considerations for wireless networked
  control systems,'' \emph{IEEE Transactions on Industrial Electronics},
  vol.~63, no.~9, pp. 5547--5557, 2016.

\bibitem{park2017wireless}
P.~Park, S.~C. Ergen, C.~Fischione, C.~Lu, and K.~H. Johansson, ``Wireless
  network design for control systems: A survey,'' \emph{IEEE Communications
  Surveys \& Tutorials}, vol.~PP, no.~99, pp. 1--36, 2017.

\bibitem{hui2012stabilizing}
H.~Cheng, Y.~Chen, W.~S. Wong, Q.~Yang, L.~Shen, and J.~Baillieul,
  ``Stabilizing and tracking control of multiple pendulum-cart systems over a
  shared wireless network,'' in \emph{Proc. CCC}, 2012.

\bibitem{demirel2014modular}
B.~Demirel, Z.~Zou, P.~Soldati, and M.~Johansson, ``Modular design of jointly
  optimal controllers and forwarding policies for wireless control,''
  \emph{IEEE Transactions on Automatic Control}, vol.~59, no.~12, pp.
  3252--3265, 2014.

\bibitem{gao2008new}
H.~Gao, T.~Chen, and J.~Lam, ``A new delay system approach to network-based
  control,'' \emph{Automatica}, vol.~44, no.~1, pp. 39--52, 2008.

\bibitem{hu2007}
S.~Hu and W.~Yan, ``Stability robustness of networked control systems with
  respect to packet loss,'' \emph{Automatica}, vol.~43, no.~7, pp. 1243--1248,
  2007.

\bibitem{tan2015stabilization}
C.~Tan, L.~Li, and H.~Zhang, ``Stabilization of networked control systems with
  both network-induced delay and packet dropout,'' \emph{Automatica}, vol.~59,
  pp. 194--199, 2015.

\bibitem{lam2007}
J.~Xiong and J.~Lam, ``Stabilization of linear systems over networks with
  bounded packet loss,'' \emph{Automatica}, vol.~42, no.~1, pp. 80--87, 2007.

\bibitem{chen2016}
H.~Chen, J.~Gao, T.~Shi, and R.~Lu, ``H$_{\infty}$ control for networked
  control systems with time delay, data packet dropout and disorder,''
  \emph{Neurocomputing}, vol. 179, pp. 211--218, 2016.

\bibitem{tan2017TAC}
C.~Tan and H.~Zhang, ``Necessary and sufficient stabilizing conditions for
  networked control systems with simultaneous transmission delay and packet
  dropout,'' \emph{IEEE Transactions on Automatic Control}, vol.~62, no.~8, pp.
  4011--4016, 2017.

\bibitem{liu2004wireless}
X.~Liu and A.~Goldsmith, ``Wireless network design for distributed control,''
  in \emph{Proc. IEEE CDC}, 2004.

\bibitem{park2011wireless}
P.~Park, J.~Ara{\'u}jo, and K.~H. Johansson, ``Wireless networked control
  system co-design,'' in \emph{Proc. IEEE ICNSC}, 2011.

\bibitem{zeyu2016autonomous}
Z.~Jiang, H.~Cheng, Z.~Zheng, X.~Zhang, X.~Nie, W.~Li, Y.~Zou, and W.~S. Wong,
  ``Autonomous formation flight of uavs: Control algorithms and field
  experiments,'' in \emph{Proc. CCC}, 2016.

\bibitem{cheng2017integrated}
C.~Tan, W.~S. Wong, and H.~Zhang, ``Integrated control over software defined
  network,'' in \emph{Proc. CCC}, 2017.

\bibitem{hou2009qos}
I.-H. Hou, V.~Borkar, and P.~R. Kumar, ``A theory of {QoS} for wireless,'' in
  \emph{Proc. IEEE INFOCOM}, 2009.

\bibitem{hou2012queueing}
I.-H. Hou and P.~R. Kumar, ``Queueing systems with hard delay constraints: a
  framework for real-time communication over unreliable wireless channels,''
  \emph{Queueing Systems}, vol.~71, no. 1-2, pp. 151--177, 2012.

\bibitem{deng2017timely}
L.~Deng, C.-C. Wang, M.~Chen, and S.~Zhao, ``Timely wireless flows with general
  traffic patterns: capacity region and scheduling algorithms,'' \emph{IEEE/ACM
  Transactions on Networking}, vol.~25, no.~6, pp. 3473--3486, 2017.

\bibitem{hou2011optimality}
I.-H. Hou, A.~Truong, S.~Chakraborty, and P.~Kumar, ``Optimality of periodwise
  static priority policies in real-time communications,'' in \emph{Proc. IEEE
  CDC}, 2011.

\end{thebibliography}

\appendix

\subsection{Proof of Theorem \ref{the:h-min-perfect}} \label{app:proof-of-h-min-perfect}
The stability condition \eqref{equ:stablity-condition-perfect-channel} is equivalent to
\be
f(h) \triangleq \sum_{i=1}^N \left[ 1 - \frac{1}{ e^{4A_ih \Delta} - e^{2A_ih\Delta} + 1} \right] - h < 0.
\ee

Instead of considering integer $h$, we consider $h \in \mathbb{R}$ for function $f(h)$.
First, it is easy to see that $f(0)=0$ and $f(h) < 0$ for all $h \ge N$.
We trace back $f(h)$ from $h=N$ to $h=0$ and find the first $h$ such that $f(h)=0$ and we denote it
as $h^*$, i.e.,
\be
h^* \triangleq \max\{h \in \mathbb{R}: 0 \le h \le N, f(h)=0\}.
\ee
Note that $h^*$ is well-defined because $f(0)=0$ and $h^* \in [0, N)$ because $f(N) < 0$.
Since $f(h)$ is continuous and $f(N) < 0$, we have that
\be
f(h) < 0, \quad \forall h \in  (h^*, N].
\ee
In addition, since $f(h) < 0$ for all $h \ge N$, we have that
\be
f(h) < 0, \quad \forall h \in  (h^*, \infty)
\ee
Then we denote
\be
h_{\min}=
\left\{
  \begin{array}{ll}
    h^*+1, & \hbox{if $h^*$ is an integer;} \\
    \lceil h^* \rceil, & \hbox{otherwise.}
  \end{array}
\right.
\ee
Since $h^* \in [0, N)$, we obtain that $h_{\min} \in [N]$.
Clearly, $f(h) < 0$ when $h \ge h_{\min}$. The proof is completed.

\subsection{Proof of Theorem \ref{thm:sequential-inequality-due-to-submodular}} \label{app:proof-of-sequential-inequality-due-to-submodular}
Define function
\be
f(\mathcal{S}) \triangleq h - \mathbb{E}[I_{\{\mathcal{S}\}}], \quad \forall \mathcal{S} \subset [N].
\ee
Then to prove Theorem \ref{thm:sequential-inequality-due-to-submodular},
we need to prove the following inequality,
\be
f(\{1\}) \ge \frac{f(\{1,2\})}{2} \ge \cdots \ge \frac{f(\{1,2,\cdots,N\})}{N}
\label{equ:app-f-inequ}
\ee

Clearly $f(\emptyset)=0$ and
\be
f(\mathcal{S}_1)=f(\mathcal{S}_2), \text{ if } |\mathcal{S}_1| = |\mathcal{S}_2|
\label{equ:app-f-equ-same-size}
\ee
due to the symmetry.
The authors in \cite{hou2011optimality} show that $f(\mathcal{S})$ is a submodular function.
Therefore, for any $\mathcal{S}_1 \subset [N], \mathcal{S}_2 \subset[N]$, we have
\be
f(\mathcal{S}_1) + f(\mathcal{S}_2) \ge f(\mathcal{S}_1 \cup \mathcal{S}_2) + f(\mathcal{S}_1 \cap \mathcal{S}_2), \label{equ:app-f-submodular-inequ-k}
\ee

Now we prove \eqref{equ:app-f-inequ} by induction.

First, when setting $\mathcal{S}_1 = \{1\}, \mathcal{S}_2 = \{2\}$ in \eqref{equ:app-f-submodular-inequ-k}, we have
\bee
 f(\{1\}) +  f(\{2\}) & \ge f(\{1\} \cup \{2\}) + f(\{1\} \cup \{2\}) \nnb \\
& = f(\{1,2\}) + f(\emptyset) = f(\{1,2\}).
\eee
In addition, due to  \eqref{equ:app-f-equ-same-size}, we have $f(\{2\}) = f(\{1\})$.
This implies that
\be
f(\{1\}) \ge \frac{f(\{1,2\})}{2}.
\ee

Second, suppose that
\be
\frac{f(\{1,2, \cdots, k-1\})}{k-1}  \ge \frac{f(\{1,2,\cdots,k\})}{k}
\label{equ:app-hypothesis}
\ee
holds for $k \ge 2$.
Now in \eqref{equ:app-f-submodular-inequ-k}, we consider two sets
$\mathcal{S}_1 = \{1,2,\cdots,k\}$ and $\mathcal{S}_2=\{2,3,\cdots,k, k+1\}$. Clearly
both $\mathcal{S}_1$ and $\mathcal{S}_2$ are of size $k$ and thus $f(\mathcal{S}_2) = f(\mathcal{S}_1) = f(\{1,2,\cdots,k\})$.
In addition, we have $\mathcal{S}_1 \cup \mathcal{S}_2 = \{1,2,\cdots, k+1\}$ and
$\mathcal{S}_1 \cap \mathcal{S}_2 = \{2,3,\cdots,k\} \triangleq \mathcal{S}_3$. Since $\mathcal{S}_3$ is
of size $k-1$, we have $f(\mathcal{S}_3) = f(\{1,2,\cdots,k-1\})$.
Thus, we have
\bee
& 2f(\{1,2,\cdots,k\}) = f(\mathcal{S}_1) + f(\mathcal{S}_2) \ge f(\mathcal{S}_1 \cup \mathcal{S}_2) + f(\mathcal{S}_3) \nnb \\
& = f(\{1,2,\cdots,k+1\}) + f(\{1,2,\cdots,k-1\}) \nnb \\
& \ge f(\{1,2,\cdots,k+1\}) + \frac{k-1}{k}f(\{1,2,\cdots,k\}), \label{equ:app-f-inequ-k+1}
\eee
where the last inequality follows from hypothesis \eqref{equ:app-hypothesis}.
Rearranging \eqref{equ:app-f-inequ-k+1}, we have
\be
\frac{f(\{1,2, \cdots, k\})}{k}  \ge \frac{f(\{1,2,\cdots,k+1\})}{k+1},
\ee
which shows that \eqref{equ:app-hypothesis} also holds for $k+1$.
Thus we complete the proof for \eqref{equ:app-f-inequ}.

\subsection{Proof of Lemma  \ref{lem:app-f(p)-increase}} \label{app:proof-of-app-f(p)-increase}

Clearly, we have
\be
X_{1 \cup \mathcal{S}} = \min\{X_1 + X_{\mathcal{S}}, h\}.
\ee
Thus,
\bee
& \mathbb{E}[X_{1 \cup \mathcal{S}}-X_{\mathcal{S}}] = \sum_{k=1}^h \mathbb{E}[X_{1 \cup \mathcal{S}}-X_{\mathcal{S}}|X_{\mathcal{S}}=k]P(X_{\mathcal{S}}=k) \nnb \\
& = \sum_{k=1}^h \mathbb{E}[\min\{X_1 + X_{\mathcal{S}}, h\}-X_{\mathcal{S}}|X_{\mathcal{S}}=k]P(X_{\mathcal{S}}=k) \nnb \\
& = \sum_{k=1}^h \mathbb{E}[\min\{X_1 + k, h\}-k]P(X_{\mathcal{S}}=k) \nnb \\
& = \sum_{k=1}^h \mathbb{E}[\min\{X_1, h-k\}]P(X_{\mathcal{S}}=k) \nnb \\
& = \sum_{k=1}^{h-1} \mathbb{E}[\min\{X_1, h-k\}]P(X_{\mathcal{S}}=k),
\eee
Note that since $1 \notin \mathcal{S}$, $P(X_{\mathcal{S}}=k)$ does not depends on $p_1$. On the contrast,
$\mathbb{E}[\min\{X_1, h-k\}]$ only depends on $p_1$.
For any $k \in \{1,2,\cdots, h-1\}$, we next compute
$\mathbb{E}[\min\{X_1, h-k\}]$ as follows,
\bee
& \mathbb{E}[\min\{X_1, h-k\}] \nnb \\
& = \sum_{x=1}^{h-k-1} x P(X_1 = x) + (h-k) P(X_1 \ge h-k) \nnb \\
& = \sum_{x=1}^{h-k-1} x (1-p_1)^{x-1} p_1 + (h-k) (1-p)^{h-k-1} \nnb \\
& = p_1 \sum_{x=1}^{h-k-1} \frac{d}{dp_1} \left[ -(1-p_1)^x \right]+ (h-k) (1-p)^{h-k-1} \nnb \\
& = -p_1 \cdot \frac{d}{dp_1} \left[ \sum_{x=1}^{h-k-1} (1-p_1)^x \right]+ (h-k) (1-p)^{h-k-1} \nnb \\
& = -p_1 \cdot \frac{d}{dp_1} \left[ \sum_{x=1}^{h-k-1} (1-p_1)^x \right]+ (h-k) (1-p)^{h-k-1} \nnb \\
& = \resizebox{.97\linewidth}{!}{$-p_1 \cdot \frac{d}{dp_1} \left[ \frac{(1-p_1)-(1-p_1)^{h-k}}{p_1} \right]+ (h-k) (1-p)^{h-k-1}$} \nnb \\
& = \resizebox{.97\linewidth}{!}{$-p_1 \cdot \left[ \frac{\left[ (h-k)(1-p_1)^{h-k-1} -1 \right] p_1 -\left[(1-p_1)-(1-p_1)^{h-k}\right]}{p_1^2} \right]$} \nnb \\
& \quad + (h-k) (1-p)^{h-k-1} \nnb \\
& = \frac{1-(1-p_1)^{h-k}}{p_1}.
\eee
Therefore,
\be
p_1  \mathbb{E}[\min\{X_1, h-k\}] = 1-(1-p_1)^{h-k},
\ee
increases as $p_1$ increase.
Thus,
\bee
f(p) & = p_1 \mathbb{E}[X_{1 \cup \mathcal{S}}-X_{\mathcal{S}}] \nnb \\
& = \sum_{k=1}^{h-1} p_1 \mathbb{E}[\min\{X_1, h-k\}]P(X_{\mathcal{S}}=k),
\eee
increases as $p_1$ increases. This completes the proof.

This completes the proof.

\subsection{Proof of Theorem~\ref{thm:stability-ipd} } \label{app:proof-of-stability-ipd}
It follows from Theorem 1 in \cite{tan2015stabilization}, when $q^i_{j}<q^i_{\max}({A}_i,h_i),~j=1,\cdots,n_i$, the scalar DARE (\ref{dare01}) has a unique positive solution $P_i>0$.
We construct the following control policy
\bee
u^i_{k} & =-\Upsilon_i^{-1}M_i \hat{x}^i_{k+1|k-1} \nnb \\
& = -\Upsilon_i^{-1}M_i \left[\bar{A}_i x^i_{k} + (1-q_i)\bar{B}_i u^i_{k-1} \right],
\label{equ:control-policy-2}
\eee
where
\be
q_i = \max_{j \in [n_i]} q_j^i < q^i_{\max}({A}_i,h_i).
\ee
For sub-system $i$, define the Lyapunov function as
\bee
&V_k(x^i_k)  \nonumber\\
=&~\mathbb{E}
\left[(P_i+\Upsilon_i^{-1}(M_i)^2)(x^i_k)^2
-\Upsilon_i^{-1}(M_i)^2x^i_k\hat{x}^i_{k|k-2} \right],
\label{lya01}
\eee
where $\hat{x}^i_{k|k-2}$ is the predicted state in frame $k$
based on the observation of the state in frame $k-1$ and the control variable in frame $k-2$ (see \eqref{equ:predicted-state}).

By utilizing the orthogonality of $x^i_k$ and $x^i_k-\hat{x}^i_{k|k-2}$, (\ref{lya01}) can be equivalently rewritten as
\begin{align*}
V_k(x^i_k)=&~\mathbb{E}\left[P_i(x^i_k)^2+\Upsilon_i^{-1}(M_i)^2 (x^i_k-\hat{x}^i_{k|k-2})^2 \right]\\
\geq & ~\mathbb{E} \left[P_i(x^i_k)^2 \right] \ge 0.
\end{align*}
Assume $k=mn_i+j$.
Then, it follows that
\begin{align*}
&V_k(x^i_k)-V_{k+1}(x^i_{k+1}) \\
=&\mathbb{E}
\left[(P_i+\Upsilon_i^{-1}(M_i)^2) (x^i_k)^2
-\Upsilon_i^{-1}(M_i)^2 x^i_k\hat{x}^i_{k|k-2} \right. \\
&-(P_i+\Upsilon_i^{-1}(M_i)^2)(\bar{A}_i x_{k}^i+ \gamma_{k-1}^i \bar{B}_i u_{k-1}^i)^2 \\
&\left. -(\bar{A}_i x_{k}^i+ \gamma_{k-1}^i \bar{B}_i u_{k-1}^i) \Upsilon_i^{-1}(M_i)^2
(\bar{A}_i x_{k}^i+  (1-q_{j}^i)\bar{B}_i u_{k-1}^i) \right]
\\
=&\mathbb{E} \left[
(x^i_k)^2+(u^i_{k-1})^2-\Upsilon_i (u_{k-1}+\Upsilon_i^{-1}M_i \hat{x}^i_{k|k-2})^2 \right.\\
&+2(q^i_{j}-q_i)\bar{A}_iP_i \bar{B}_i(\hat{x}^i_{k|k-2}) u^i_{k-1} \\
&+((1-q_i)^2-(1-q^i_{j})^2)P_i (\bar{B}_i)^2 (u^i_{k-1})^2 \\
&\left. + (q_{i}(1-q_{i})-q^i_{j}(1-q^i_{j}))({P}_i+\Upsilon_i^{-1}(M_i)^2) (\bar{B}_i)^2 (u^i_{k-1})^2 \right].
\end{align*}
Applying the control policy \eqref{equ:control-policy-2} to the above equation, we obtain
\begin{equation*}\label{eq006}\begin{aligned}
&V_k(x^i_k)-V_{k+1}(x^i_{k+1}) \\
=&\mathbb{E} \big[(x^i_k)^2+ ({u}^i_{k-1})^2  \big]+\frac{q_{i}-q^i_{j}}{1-q_{i}} \mathbb{E} \big[ \Upsilon_i^{-1} (M_i )^2(\hat{x}_{k|k-2} )^2 \big]\\
&+\mathbb{E} \Big[ \big(\frac{q_{i}-q^i_{j} }{1-q_{i}}+(q_{i}-q^i_{j})(1-q^i_{j} ) \Upsilon_i^{-1} (M_i)^2 (\bar{B}_i)^2 \big) \\
&\times (\Upsilon_i^{-1}M_i
\hat{x}_{k|k-2})^2   \Big],
\end{aligned}\end{equation*}
which implies that $V_k(x_k)-V_{k+1}(x_{k+1})>0$ for $0<q^i_{j}\leq q_i<1$ when state $x_k$ is not zero for finite $k$.
Therefore, based on Lyapunov stability theory, sub-system $i$ is stabilizable in the mean square sense.

\begin{IEEEbiography}[{\includegraphics[width=1in,height=1.25in,clip,keepaspectratio]{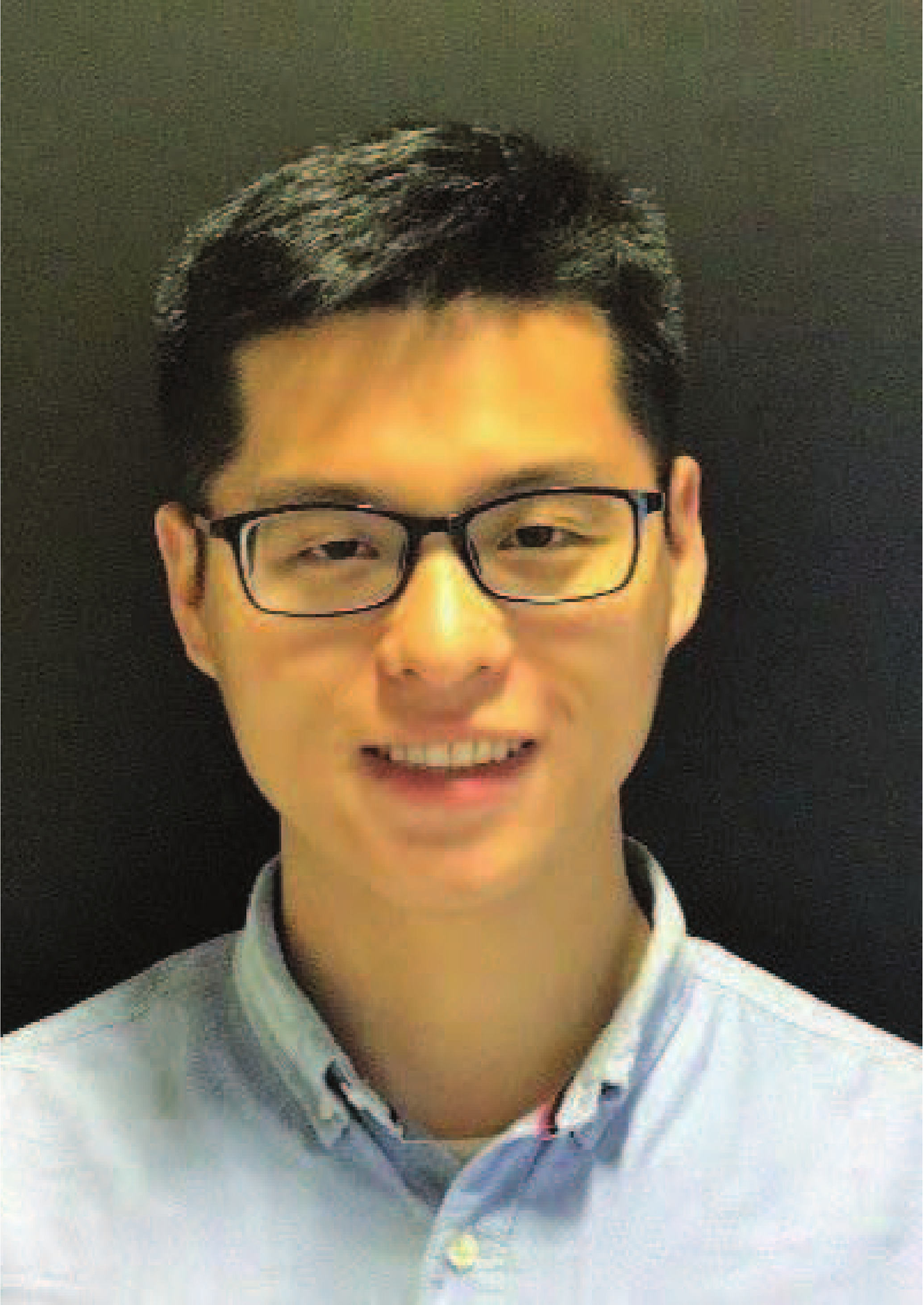}}]{Lei Deng}(M'17)
 received the B.Eng. degree from the Department of Electronic Engineering, Shanghai Jiao Tong University, Shanghai, China, in 2012, and the Ph.D. degree from the Department of Information Engineering, The Chinese University of Hong Kong, Hong Kong, in 2017. In 2015, he was a Visiting Scholar with the School of Electrical and Computer Engineering, Purdue University, West Lafayette, IN, USA. He is now an assistant professor in School of Electrical Engineering \& Intelligentization, Dongguan University of Technology. His research interests are timely network communications, intelligent transportation system, and spectral-energy efficiency in wireless networks.
\end{IEEEbiography}

\begin{IEEEbiography}[{\includegraphics[width=1in,height=1.25in,clip,keepaspectratio]{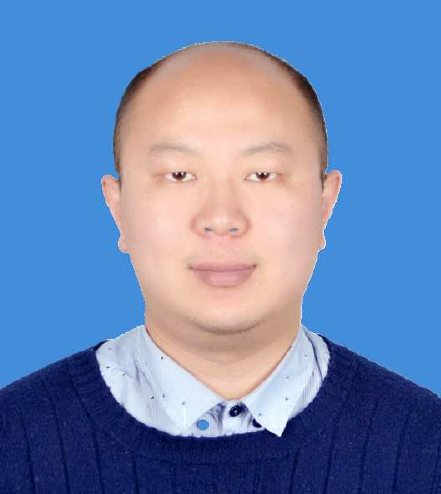}}]{Cheng Tan}(M'18)
received the B.S. degree and M.S. degree from School of Information Science and Engineering, Shandong University of Science and Technology, Qingdao, China, in 2010 and 2012, and the Ph.D. degree in School of Control Science and Engineering, Shandong University, Jinan, China in 2016.
He is now a Post-Doctoral Fellow with the Department of Information Engineering, The Chinese University of Hong Kong, Shatin, N. T., Hong Kong.
He is also an assistant professor in College of Engineering, Qufu Normal University.
His research interests include networked control system, stochastic control, time-delay system, and optimization control.
\end{IEEEbiography}

\begin{IEEEbiography}[{\includegraphics[width=1in,height=1.25in,clip,keepaspectratio]{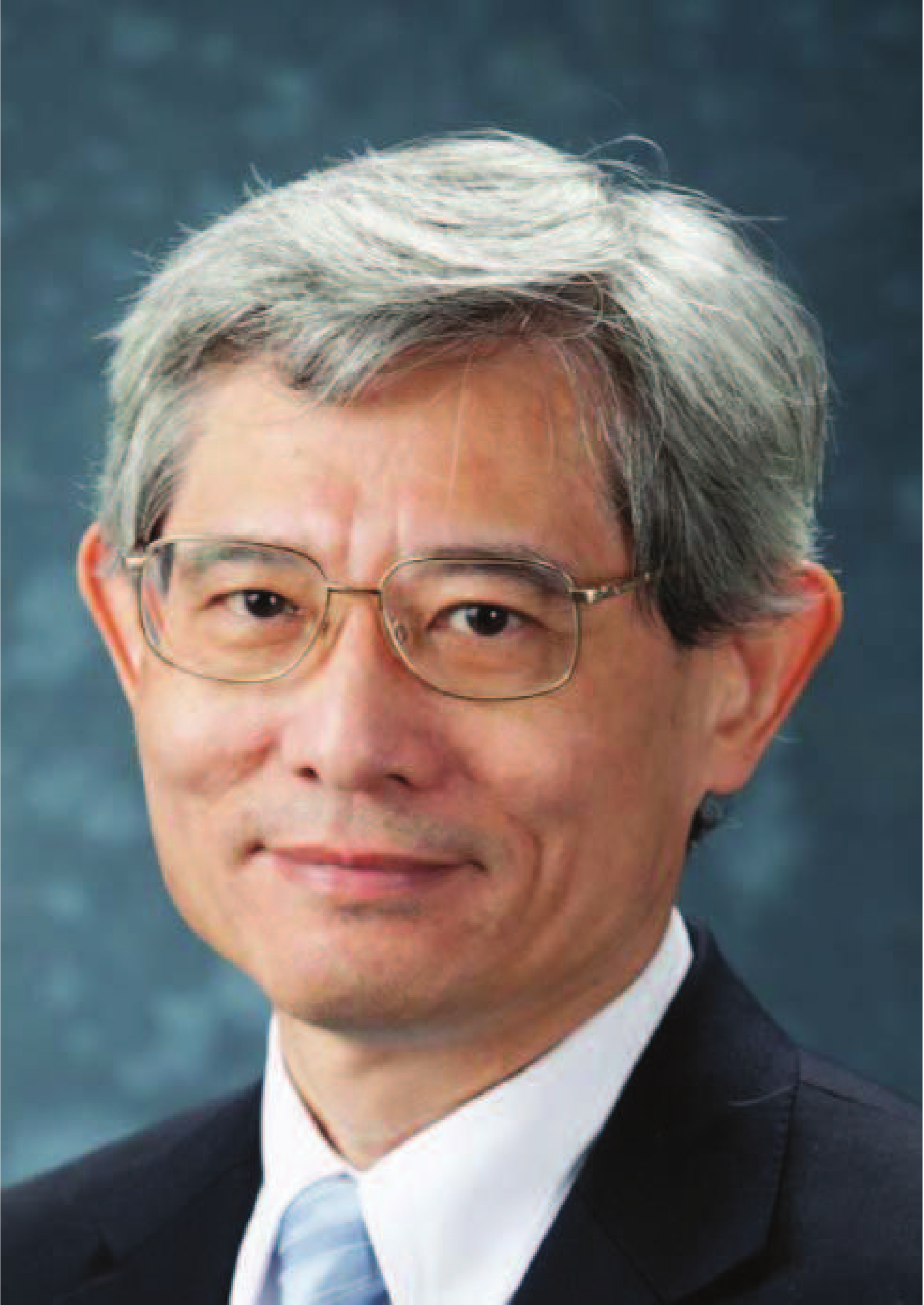}}]{Wing Shing Wong}(M'81--SM'90--F'02) received a combined master and bachelor degree from Yale University and M.S. and Ph.D. degrees from Harvard University.
He worked for the AT\&T Bell Laboratories from 1982 until he joined the Chinese University of Hong Kong in 1992, where he is now Choh-Ming Li Research Professor of Information Engineering. He was the Chairman of the Department of Information Engineering from 1995 to 2003 and the Dean of the Graduate School from 2005 to 2014.  He served as Science Advisor at the Innovation and Technology Commission of the HKSAR government from 2003 to 2005.  He has participated in a variety of research projects on topics ranging from mobile communication, networked control to network control.
\end{IEEEbiography}

\vspace{-1cm}

\end{document}